\documentstyle[aps,eqsecnum,preprint,amsfonts,amssymb,epsfig]{revtex}

\draft \tighten

\title{Quasiclassical theory of superconductivity:\\
a multiple interface geometry}
\author{A. Shelankov$^{*}$ and M. Ozana}

\address{
Department of Theoretical Physics, Ume{\aa} University, 901 87
Ume{\aa}, Sweden 
}

\let\Samizdat = \relax

\newcommand{\Format}[2]{%
\bbox{{\cal F}_{R}\left[ \rule[-#1ex]{0ex}{0ex}\right.}%
#2%
\bbox{\left. \rule[-#1ex]{0ex}{0ex}\right]}%
}

\newcommand{\gR}{\hat{g}^{R}}

\def\baroverletter#1{\setbox1=\hbox{$#1$}
  \dimen1=\ht1
    \advance\dimen1 by 1pt
  \dimen2=\wd1
    \advance\dimen2 by -1pt
  \rlap{\hspace{.5pt}\rule[\dimen1]
              {\dimen2}{.35pt}}\box1} 

\def\ops{{\baroverletter\psi}}

\def\ofi{{\baroverletter\phi}}

\newcommand{\psie}{\psi_{e}}
\newcommand{\psih}{\psi_{h}}

\newcommand{\Hrow}[2]{\left(
\begin{array}{lcr}
#1   &,& #2
\end{array}
\right)
}

\newcommand{\Hcolumn}[2]{\left(
\begin{array}{c}
#1 \\
#2
\end{array}
\right)
}

\begin{document}

\maketitle

\begin{abstract}
A new method which allows one to study multiple coherent
reflection/transmissions by partially transparent interfaces, (e.g.,
in multi-layer mesoscopic structures or grain boundaries in
high-Tc's), in the framework of the quasiclassical theory of
superconductivity is suggested. It is argued that in the presence of
interfaces, a straight-line trajectory transforms to a simple
connected 1-dimensional tree (graph) with knots, i.e. the points where
the interface scattering events occur and pieces of the trajectories
are coupled.  For the 2-component trajectory "wave function" which
factorizes the Gor'kov matrix Green's function, a linear boundary
condition on the knot is formulated for an arbitrary interface,
specular or diffusive (in the many channel model).  From the new
boundary condition, we derive: (i) the excitation scattering amplitude
for the multi-channel Andreev/ordinary reflection/transmission
processes; (ii) the boundary conditions for the Riccati equation;
(iii) the transfer matrix which couples the trajectory Green's
function before and after the interface scattering.  To show the usage
of the method, the cases of a film separated from a bulk
superconductor by a partially transparent interface, and a SIS'
sandwich with finite thickness layers, are considered.  The electric
current response to the vector potential (the superfluid density
$\rho_s$) with the $\pi $ phase difference in S and S' is calculated
for the sandwich.  It is shown that the model is very sensitive to
imperfection of the SS' interface: the low temperature response being
paramagnetic ($\rho_s <0$ ) in the ideal system case, changes its sign
and becomes diamagnetic ($\rho_s > 0$) when the probability of
reflection is as low as a few percent.  \end{abstract}

\bigskip




\section{Introduction}\label{intro}

Many important properties of superconductors are related to surfaces
and interfaces, the Josephson and proximity effects being well-known
examples.  In recent years, new rich surface physics has been found in
high-T$_c$ oxides after the identification of the d-symmetry of the
order parameter.  On the theoretical side, studying an interface poses
certain problems: The method of the quasiclassical Green's functions
\cite{Eil68,LarOvc,Sch81,SerRai83} (for a recent review see
\cite{karlsruhe} )which is the main tool in the superconductivity
theory, cannot be directly applied here since the quasiclassical
condition is violated by fast change of the potentials on the atomic
distances in the vicinity of the interface.  As shown by Zaitsev
\cite{Zai84}, the abrupt changes at a specular partially transparent
interface can be incorporated into a boundary condition for the
quasiclassical Green's functions; the condition is a third order
equation for the matrix Green's function near the interface.  Various
forms of the boundary condition have been discussed in more recent
papers \cite{AshAoyHar89,Nag98,Yip97}.  New difficulties arise when
one attempts to describe the coherent reflection/transmission by many
interfaces, e.g.  in a multi-layer mesoscopic structures or grain
boundaries network in high-Tc's . In this case, Zaitsev's third order
boundary condition must be satisfied on each interface, and one
encounters the problem of solving a system of cubic matrix equations.
It is not obvious that a solution to the system of equations exists
and is unique if it exists.  Moreover, some authors
\cite{AshAoyHar89,Nag98} doubt the very applicability of the
quasiclassical scheme in the many interface geometry: They argue that
the quasiclassical normalization, which is a vital part of the
quasiclassical scheme, is not possible in a double layer system with
partially reflective interface.

The purpose of the present paper is to re-examine the theory of the
interface in the quasiclassical description of superconductivity. A
new scheme which allows one to incorporate specular as well as
diffusive interface(s) into the quasiclassical theory is suggested.
To make the presentation self-contained, we start with a short
introduction to the quasiclassical theory of superconductivity.

As first shown by Bardeen, Cooper, and Schrieffer (BCS) \cite{deG89},
the phenomenon of superconductivity can be understood in the framework
of a mean-field type scheme where the Cooper correlations are
introduced through the pair potential $\Delta $ (generally, a function
of the momentum $\bbox{p}$) which is related to electron-electron
interaction by a self-consistency condition.  The mean field $\Delta $
may be introduced directly as a kind of Hartree-type potential, or it
can be derived in the framework of a more sophisticated Eliashberg
theory where the pair potential comes as the anomalous self-energy in
the Gor'kov equations for the Green's function.  This truly
microscopic approach allows one to perform all the normalizations in
the spirit of the Landau theory of Fermi liquid and to consider
superconductors with a strong coupling (see Seren and Rainer
\cite{SerRai83} and references therein.)

Whatever the method of derivation, the Gor'kov equation for the matrix
Green's function gives the basis for studying the BCS-type
superconductivity. The quasiclassical theory of superconductivity
offers an approximate simplified scheme of solving the Gor'kov
equation.  To clarify physics behind the approximations, we analyze
first the Bogoliubov - de Gennes equation \cite{deG89} that is the
effective ``Schr\"odinger equation'' corresponding to the Gor'kov
equation (in the weak coupling limit).

It is well--known that Cooper's pairing in the superconducting state
is conveniently described in the language of the electron-hole
coherence.  On the mean field level, the ground as well as an excited
states of the system are products of single particle states, each of
them a quantum superposition of electron and hole.  The electron,
$\psie$, and hole, $\psih$, amplitudes in the superposition comprise
the 2-component single particle wave function, $\Psi (\bbox{r},t) =
{\psie \choose \psih}$.  It obeys the Bogoliubov - de Gennes equation
\cite{deG89},
\begin{equation}
i \hbar  {\partial\over{\partial t}}
\left(
\begin{array}{c}
\psie       \\
\psih
\end{array}
\right)
=
\left(
\begin{array}{cc}
\xi( \hat{\bbox{p}}- {e\over c} \bbox{A}) +U   & \Delta   \\
\Delta^*   &-\xi( \hat{\bbox{p}}+ {e\over c} \bbox{A})-U
\end{array}
\right)
\left(
\begin{array}{c}
\psie       \\
\psih
\end{array}
\right) 
\label{Ooa}
\end{equation}
where $\xi (\bbox{p})= \epsilon(\bbox{p})-\mu$, $\epsilon (\bbox{p})$
and $\mu $ being the electron band energy and the chemical potential,
respectively, $\bbox{A}$ is the magnetic vector potential;
$U(\bbox{r})$ is the potential energy.  The pairing potential $\Delta
$, and, in principle, all other potentials must be found
self-consistently.

For future needs we note that in the vicinity of the the Fermi surface
$\xi (\bbox{p}_{F})=0$, the electron (hole) with the momentum
$\bbox{p}\approx \bbox{p}_{F}$ moves with the Fermi velocity
$\bbox{v}= +(-) {\partial \bbox{\xi }\over{\partial
\bbox{p}_{F}}}|_{\xi=0}$.  The particle energy is close to the Fermi
energy $E_{F}\sim v p_{F}$, and the de-Broglie wave length is
$\lambdabar_{F}$ is of order of $\lambdabar_{F}\sim \hbar /p_{F}$,
$p_{F}$ being a typical momentum on the Fermi surface.

In the superconductors which are good metals in the normal state, the
potentials are semiclassical (excluding interfaces and disorder which
are discussed later) {\it i.e.} they are slowly varying functions of
the coordinate on the scale of the wave length $\lambdabar_{F}$.
Indeed, the pair potential $\Delta $ changes at the coherence length
$\xi_{0} \sim \hbar v/ \Delta$, and one estimates the ratio
$\lambdabar_{F}/ \xi_{0}$ as $\lambdabar_{F}/ \xi_{0} \sim \Delta
/E_{F}$.  Also, the validity of a semiclassical treatment of magnetic
field $B$ requires that $\lambdabar_{F}<<l_{B} $, $l_{B}$ being the
magnetic length, $l_{B}= \sqrt{\Phi_{0}/B}\; , \; \Phi_{0}= hc/2e
$. Since superconductivity exists only at $B< B_{c2}\sim \Phi_{0}/
\xi_{0}^{2}$ , the ratio $\lambdabar_{F}/l_{B}$ never exceeds $\Delta
/ E_{F}$.  Seeing that $\Delta \sim T_{c} $, the semiclassical
conditions $\lambdabar_{F}/ \xi_{0}\;,\; \lambdabar_{F}/l_{B}\ll 1$
are equivalent to the requirement that $ T_c /E_{F} << 1$.  In
accordance with the Landau theory of Fermi liquid, this condition is
always is satisfied if the normal state is metallic.

Most of the physical effects in metals and superconductors (the Hall
and thermoelectric effects being notable exceptions) can be described
in the simplest approximation where all the corrections of order
$T/E_{F}\sim T_{c}/E_{F}$ are neglected {\it i.e.}  in the limit
$T_{c}/E_{F} \rightarrow 0$.  This is the approximation where the
quasiclassical theory of superconductivity is valid
\cite{SerRai83,She80c}.  

Since $ {T_{c}\over E_{F}} \sim {\hbar \over p_{F}\xi_{0}}$, the limit is
equivalent to $\hbar \rightarrow 0$ or large mass $m \sim
p_{F}/v\rightarrow \infty $.  In this limit of quantum mechanics of
noninteracting particles, wave packets do not suffer quantum
broadening and dynamics becomes completely classical: The particle
moves along a trajectory, position $\bbox{r}(t)$ and momentum
$\bbox{p}(t)$ being well defined.  Below we analyse how the
electrons-hole coherence  in the superconducting
state changes the situation.

First, we consider in more detail the classical dynamics of the
electron and hole separately.  The Bogoliubov - de Gennes equation
where we put $\Delta =0$ for the moment, reads
\begin{equation}
i \hbar {\partial \psie\over{\partial t}}= 
\left[\xi \left(\hat{\bbox{p}}- e \bbox{A}\right) + U \right]\psie
\;\; , \;\; 
i \hbar {\partial \psih\over{\partial t}}= 
-\left[\xi \left(\hat{\bbox{p}}+ e \bbox{A}\right) +  U \right]\psih
\label{Oka}
\end{equation}
The two equations transform into each other after the substitution $t
\rightarrow -t$ and $\bbox{A} \rightarrow - \bbox{A}$. This means that
given a solution $\psie(\bbox{r},t |\{\bbox{A}\})$ corresponding to
the vector potential $\bbox{A}$, the function $\psie(\bbox{r},-t
|\{-\bbox{A}\})$ solves the equation for $\psih$ in the vector
potential $\bbox{A}$.
Therefore, 
\begin{equation}
 \psih(\bbox{r},t
|\{\bbox{A}\})= \psie(\bbox{r},-t |\{-\bbox{A}\})\;,
\label{Oma}
\end{equation}
provided $\psie(\bbox{r},t=0)=\psih(\bbox{r},t=0)$ \cite{e-h}.

If $\hbar \rightarrow 0$, the centre of electron or hole wave packets
moves in the $\bbox{r}-\bbox{p}$ space along the trajectory specified
by the coordinate $\bbox{r}_{e,h}(t)$ and momentum $\bbox{p}_{e,h}(t)$
as a function of time $t$. The relation between electron- and
hole-trajectories can be expressed in the following way.

Let $\bbox{r}_{e(h)}(t | \{\bbox{b}\})$ together with  $\bbox{p}_{e(h)}(t |
\{\bbox{b}\})$ be the trajectory of the electron (hole) in the magnetic field
 $\bbox{b}= \bbox{\text{rot }} \bbox{A}$. 
From Eq.(\ref{Oma}) one can conclude that the corresponding classical
dynamics of electrons and holes are related to each other in the
following way:
\begin{equation}
\bbox{r}_{h}(t | \{\bbox{b}\})=
\bbox{r}_{e}(-t | \{-\bbox{b}\})
\;\; , \;\;  
\bbox{p}_{h}(t | \{\bbox{b}\})=
\bbox{p}_{e}(-t | \{-\bbox{b}\}
\label{Ona}
\end{equation}
provided 
the electron and hole trajectories pass through
the same point
 $\bbox{r}_{e}=\bbox{r}_{h}= \bbox{r}_{0}$
 and
 $\bbox{p}_{e}=\bbox{p}_{h}= \bbox{p}_{0}$ at $t=0$. 

One sees from here that if the magnetic field is absent, $\bbox{b}=0$,
or its influence on the classical dynamics is negligible, then
\begin{equation}
\bbox{r}_{h}(t)=
\bbox{r}_{e}(-t)
\;\; , \;\;  
\bbox{p}_{h}(t)=
\bbox{p}_{e}(-t) \; ,
\label{Orc}
\end{equation}
that is the electron and hole move in opposite directions along the
{\it same} line (path) in the $\bbox{r}-\bbox{p}$ space.  However, to
the extent the magnetic field influences the orbits, the electron and
hole paths are {\it different} \cite{Larmor}.  
(Obviously, the role of the magnetic field may play any perturbation
violating the time reversal symmetry.)

Now we are in position to analyse how the electron-hole mixing ({\it
i.e.} $\Delta \neq 0$ in Eq.(\ref{Ooa})) changes propagation of the
wave packets.  Consider a wave packet which is initially purely
electronic ($\psih =0 \,,\, t=0$), and assume for the moment that
Eq.(\ref{Orc}) is valid.  The electron moves classically on a
trajectory in the $\bbox{r}-\bbox{p}$ space, and provides a source,
$\Delta^{*}\psie$, in the equation for $\psih$ (see Eq.(\ref{Ooa}))
generating a hole wave.  Since $\Delta $ is a slowly varying field,
the source $\Delta^{*}(\bbox{r})\psie (\bbox{r},t)$ and
$\psie(\bbox{r},t)$ are peaked at the same point of the
$\bbox{r}-\bbox{p}$ space.  In other words, the hole is created at the
point of the current position of the electron and with the
instantaneous electron momentum.  Then, by virtue of Eq.(\ref{Orc}),
the secondary hole moves backwards along the path of the primary
electron. In turn, the hole creates new electrons which move along the
same path {\it etc}. It is very important that the multiple processes
of the electron-hole conversion keep the packet on a line in the
$\bbox{r}-\bbox{p}$ space which is nothing but the classical
trajectory.  However, the width of the packet {\it along} the
trajectory grows linearly in time $\propto vt$ (at times $t > \hbar/
\Delta $) due to the reverse of the velocity under the
electron$\,\leftrightarrow\,$hole conversion processes.

One sees that, the wave packet in a superconductor experiences
broadening even in the limit $\hbar \rightarrow 0$, and, therefore, a
quantum description is unavoidable.  Nevertheless, the notion of the
classical trajectory as a line in the $\bbox{r}-\bbox{p}$ space remains
meaningful because the quantum broadening occurs only along the line.
Ultimately, this important feature is due to the time reversal
symmetry. It holds to the extent Eq.(\ref{Orc}) is accurate {\it i.e.}
when one can neglect the magnetic Lorentz force in the classical
dynamics.

Note the peculiar role of a magnetic field: the difference in the
magnetic bending of electron and hole trajectories results in the
broadening of the coherent electron-hole wave packet in the direction
transverse to the classical trajectory. At energies $\sim \Delta $
where the electron and hole components have comparable weight, the
significance of the Lorentz force can be estimated \cite{lorentz} from
the ratio ${\hbar \omega_{c}\over \Delta }\lesssim \Delta /E_{F}$
where $\omega_{c} = |eB/mc|$ is the cyclotron frequency. Since ${\hbar
\omega_{c}\over \Delta }\propto {1\over m}$, one can consistently
neglect the Lorentz since the quasiclassical theory is effectively a
theory of infinitely heavy particles, $m \rightarrow \infty $ as
discussed before.  It seems that in general case the Lorentz force can
be incorporated in a theory of superconductivity only by a full
quantum approach (see, however, Kopnin's quasiclassical theory of the
Hall effect \cite{Kop96}).  Sometimes, the magnetic broadening may
turn out to be non-crucial, e.g. in a spatially homogeneous case, and
then certain simplifications may be possible (see
e.g. \cite{GorSch98}).

A more formal and rigourous analysis of electron-hole coherence on
classical trajectories can be done using a method  first
suggested by Andreev \cite{And64}.  The stationary state wave function
is written as $\Psi (\bbox{r},t ) = \psi (\bbox{r}) e^{{i\over{\hbar}}
\bbox{p}_{F}\bbox{\cdot r}} e^{{-i\over{\hbar}}E t}$ where $\psi
(\bbox{r})$ is a slowly varying function (provided $|E|\ll E_{F}$).
Plugging $\Psi (\bbox{r},t )$ into the Bogoliubov - de Gennes equation
Eq.(\ref{Ooa}), and using the approximation
\[
e^{-{i\over{\hbar}} \bbox{p}_{F}\bbox{\cdot r}}
\xi( \hat{\bbox{p}}- {e\over c} \bbox{A})
e^{{i\over{\hbar}} \bbox{p}_{F}\bbox{\cdot r}}
\approx \bbox{v\cdot}\left( {\hbar\over i} \bbox{\nabla } - {e\over c}\bbox{A} \right)
\] 
where the small terms of order $\left(\lambdabar_{F}\bbox{\nabla }
\right)^{2}$ are neglected, one gets the Andreev equation.
Rearranging terms, the Andreev equation may be written in the following form
\begin{equation}
\left(
i \hbar  \bbox{v\cdot \nabla }
+
\left(
\begin{array}{lr}
E - \bbox{v\cdot p}_{s}&  \Delta\\
-\Delta^{*}& -E + \bbox{v\cdot p}_{s}
\end{array}
\right) \right)
\left(
\begin{array}{c}
   \psie\\
  - \psih
\end{array}
\right) =0
\label{5db}
\end{equation}
where $\bbox{v}$ is the velocity at the point $\bbox{p}_{F}$ of the
Fermi surface, $\bbox{p}_{s}$ denotes $\bbox{p}_{s}= - {e\over
c}\bbox{A}$ and for simplicity $U=0$ (as it usually the case because
of the efficient screening). The most important feature here is that
the derivative $\bbox{v\cdot \nabla}$ couples the value of the wave
function only on straight lines in the direction of the velocity
$\bbox{v}$; the lines are the classical trajectories when $U=0$
\cite{traj}.  In this approximation, the quantum coherence exists only
along the classical trajectories without any coupling between
neighbouring paths. These properties are in agreement with the
qualitative picture of the wave packet spreading along the classical
trajectory, discussed previously.  One may call the envelope function
$\psi$ in Eq.(\ref{5db}) the wave function on the classical
trajectory.

After this short review of the quasiclassical approximation, our next
step is to include the interface into the scheme.  In this
introductory part of the paper, we present main ideas using the
language of the wave functions on classical trajectory; a more general
approach of 2-point trajectory Green's function is presented in
Section \ref{green}.

The reflection/transmission on an isolated interface (a specular one,
to begin with) mixes together semi-infinite pieces of classical
trajectories (see Fig.\ref{cross}).  Each of the pieces is
characterized by the Fermi surface momentum $\bbox{p}_{F}$, and the
corresponding  velocity is $\bbox{v}$; the arrows indicate the direction
of the velocity.   On pieces of trajectories 1 and 2 the
velocity is directed towards the interface, and we call them
in-coming trajectories (or channels); correspondingly, 1'
and 2' are outgoing (pieces of) trajectories. Throughout the paper,
the out-going ``channels'', alias for ``trajectory'', are marked by ``prime''.

Note that the in/out classification of the trajectories in accordance
with the direction of the Fermi surface velocity is unique but it is arbitrary
because the electron and hole belonging to same channel have the opposite
directions of their velocities.  
For instance, the electron coming to the interface on via e.g. the
channel 2 (see Fig. \ref{cross}) may go away as the electron on
trajectories 1' and 2'  as well as a hole along nominally in-coming
trajectory 1.

We will call ``knot'' the region inside of which scattering occurs and
the pieces of the classical trajectories get ``tied'' together on the
interface \cite{full-cross}.  Usually the typical thickness of the
interface region is of atomic scale, and only the wave function in the
outer region is of interest. Then, on the quasiclassical level of
accuracy, the interface (the knot) can be described by the scattering
matrix \cite{She80b}

In general, the knot may tie together arbitrary number, $N$, of
ballistic in-channel to the same number $N$ of the out-channels.  For
a specular interface, number of channels $N$ equals to 2, and rough
interfaces may be modelled by knots with $N>2$.

The waves generated by a source, e.g. on path 1 in Fig.\ref{cross},
spread to all other paths 1', 2, and 2' coupled by the knot.  In the
presence of an interface, the wave function on trajectory remains a
valid concept if one interprets the notion of trajectory in a broader
sense as a set of the points on all the ballistic paths coupled by the
knot.  For instance, in Fig.\ref{cross}, one understands paths 1, 2,
1', 2' as the parts of a single geometrical object, which we also call a
``trajectory''.  The spatial argument of the wave function will span
the generalized trajectory. 
Similar constructions are known in the
literature: see, e.g. Ref.\cite{KotSmi97} where the Schr\"odinger
equation is solved on graphs (networks).

The case of many interfaces requires some preliminary remarks.
Consider as an example a two layer system Fig.\ref{loop1}.  {\em If}
the layers are of the same thickness and the reflections are exactly
specular, the two outgoing path 1' and 2' meet together again on the
upper knot, forming a loop {\it i.e.}  a pair of interfering paths.
This causes a major difficulty for the quasiclassical theory: Indeed,
the envelope function $\psi $ obeying the Andreev equation is
introduced when the phase factor $e^{i p_{F}{\cal L}/ \hbar }$, ${\cal
L}$ being the distance along the path, is singled out of the full wave
function.  When loops are present and there is more than one path
connecting any 2 points, the distance ${\cal L}$ is ill-defined, and
the procedure of constructing the envelope $\psi $ becomes non-unique
and dubious. Besides, the interference phase factors like $e^{i
p_{F}({\cal L}_{1}- {\cal L}_{2})/ \hbar }$, ${\cal L}_{1,2}$ being
the lengths of the interfering paths, crucially sensitive to the value
of $p_{F}$ and cannot be found in the quasiclassical limit where $
\hbar / p_{F}= 0$.

To overcome the difficulty we note the following: The interference
leads to Fabri-Perot type geometric resonances and related
fluctuations of various physical quantities, perhaps locally strong.
However, in the limit $\hbar \rightarrow 0$, the resonances are {\em
close} to each other in the configuration space, and, therefore, the
fluctuations are expected to be effectively averaged out when one
calculates observables: The latter are given by certain integrals and
thus are sensitive mainly to coarse-grain features in the
configuration space.

Further, the coarse-grain features (like e.g. the angular-resolved
local density of states averaged in small volume ($\gg
\lambdabar_{F}^{3}$)) or small interval of directions) are more than
likely not perceptive to small variations of geometry shifting the
positions of the resonances.  Hence, it seems plausible to assume that
the coarse-grain structure can be faithfully reproduced if one
introduces ``virtual roughness'', which is small ($\ll \xi_{0} $) and
not noticeable quasiclassically, and performs averaging with respect
to the roughness (kind of ergodic hypothesis).  In other words, on the
course-grain level, an ideal surface is expected to be
indistinguishable from a ``virtually rough'' {\it i.e.} a random
surface with roughness $W$ (see Fig.\ref{loop2}) small on the typical
quasiclassical scale, $W\ll \xi_{0} $.

For a rough surface, the picture of trajectories shown in
Fig.\ref{loop1} {\em almost} never occurs: In the quasiclassical
approximation, the trajectories are lines with zero ($\sim
\lambdabar_{F}$) width, and the condition that the trajectories 1' and
2' cross each other again exactly at the interface (up to $\sim
\lambdabar_{F}$), is very restrictive. For this, the surfaces must be
strictly parallel and the reflections 1'$\rightarrow $ 3 and
2'$\rightarrow $4 must be specular (identical) with high precision.
Qualitatively, the argument here is the same as in the billiard theory
where closed orbits are known to be rare exceptions.  As long as the
loops are absent, solutions to the Andreev equation, vary smoothly
when parameters of the trajectory (e.g. its direction) or the surface
roughness are changed and have certain limit when the virtual
roughness tends to zero. Hence, the averaging with respect to the
virtual roughness is trivial: it amounts to neglecting it in any
calculation provided the topology of the trajectories is
single-connected.  The virtual roughness (tending to zero) is needed
here only as a mean to eliminate the geometric resonances which are
not of interest because they are not seen on the coarse-grain level of
description.  (Another line of reasoning could be to say that any real
sample is always microscopically rough so that loops are statistically
impossible).

By these arguments, one comes to the important conclusion that due to
the virtual (or real) roughness the paths tied together by a knot do
not show any further correlations and do not ({\em typically}) meet
each other on other knots.  This seems to be an analog to the impurity
averaging.  Effectively, it allows one to average over the Fermi wave
length scale from the very beginning.

Uncorrelated multiple collisions with interfaces transform a ballistic
trajectory into a tree-like geometrical object.  To give a general
idea of what we mean by a tree, the topological structure of one of
the possible trees with N=2,3 knots is shown in Fig.\ref{3-Fig}.  The
tree corresponding to a real physical situation will be presented
later.

The main feature of the tree-like trajectory is its one-dimensional
character, the property which can equivalently formulated as (i) there
is no loops or interfering paths; (ii) there is only one path
connecting any two points of the tree; (iii) the cut of any line
produces two disconnected pieces.

Since the tree is effectively 1-dimensional, one is able to repeat
Andreev's procedure on a tree-like trajectory defining the slowly
varying envelope wave $\psi (\bbox{r})$ by the formula $\Psi
(\bbox{r}) = \psi(\bbox{r}) e^{ip_{F}{\cal L(\bbox{\bbox{r}})}}$,
where $\bbox{r}$ spans the points on the tree, and ${\cal
L}(\bbox{r})$ is the coordinate along the tree counted off a point.
In between knots, the Andreev equation Eq.(\ref{5db}) is valid and the
values of the wave function on a knot are coupled by the scattering
S-matrix (see Section \ref{knot}).

The purpose of present paper is to extend the existing quasiclassical
Green's function theory of superconductivity to the case of
multi-interface geometry.  In essence, the standard quasiclassical
(``$\xi$-integrated'') theory of superconductivity is the Green's
function version of the Andreev equation: Again, the quantum coherence
of the electron and hole residing on the same trajectory is taken into
full consideration whereas the coherence between particles occupying
different trajectories is neglected.  The paths are coupled to each
other only by the self-consistent effective potentials like various
self-energies (impurity, phonon) and the pair potential $\Delta $.
The Green function technique has obvious advantages for one is able to
perform the disorder averaging, include the inelastic scattering and
the strong-coupling effects {\it etc}.

Although the potential due to crystal imperfections like impurities is
far not slowly varying, this does not invalidates the quasiclassical
scheme if one is interested only in the disorder averaged properties.
It is well-known that the disorder averaging amounts to the impurity
self-energy term in the Gor'kov equation which effect is similar to
that of the potential energy.  The self-energy varies on the same
spatial scale as other self-consistent potential and as such does not
violate classicality.  Of course, the imaginary part of the
self-energy must be small so that the mean free path $l$ is large,
$l\gg \lambdabar_{F}$.  The quantum localization corrections
controlled by the parameter $\hbar / p_{F}l \ll 1$ are ignored, which
again is consistent with the limit $\hbar \rightarrow 0$ or $p_{F}
\rightarrow \infty $ accepted in the quasiclassical theory.

We use the version of the quasiclassical theory \cite{She80a,She85}
where the main object is the 2-point Green's function on classical
trajectories. In our opinion, this approach is most adequate to the
above physical picture of the electron-hole phase coherence spreading
along classical trajectories. As has already been discussed, in the
many-interface geometry the classical trajectory becomes
tree-like. Accordingly, the arguments of the 2-point Green's function
are points on a tree.  In the present paper, we restrict ourselves to
the stationary case, and our main concern is the retarded Green's
function of the Keldysh technique.

The paper is organized as follows.  In Sect.\ref{green}, we review the
quasiclassical theory in the formulation based on the 2-point Green's
function.  The connection to the standard technique is discussed in
Sect.\ref{1point}.  In Sect.\ref{motion}, we briefly show the
connection to the Riccati equation technique \cite{SchMak95,Sch98}, as
well as suggest a general method for the case of a periodic
potential. In Sect.\ref{knot}, we derive the boundary conditions for
the Green's function on the knot (interface) with arbitrary number of
channels. In Sect.\ref{andreev}, a solution to the multi-channel
problem of the Andreev reflection as well as the bound states, is
given. In Sect.\ref{match}, we derive the interface boundary condition
for the Riccati equation. In Sect.\ref{transs}, the boundary condition
for the Green's function in terms of the transfer matrix is derived.
In Sect.\ref{layer}, we show the usage of the general approach
applying the theory for studying simple examples: (i) a film separated
by a partially transparent interface from of a bulk material
superconductor; (ii) two layers of a finite thickness.  Motivated by
the recent theory of the paramagnetic effect \cite{FauBelBla99}, we
pay most attention to the case when the phases of the order parameter
in the two superconductors differ in $\pi $; numerical data for the
density of states and superfluid density are presented.  The results
are summarized in Sect.\ref{concl}.  Details of the calculations are
collected in the Appendices. In the rest of the paper, $\hbar =1$.

\section{Trajectory 2-point Green's function  }\label{green}

A convenient starting point is the formulation of the quasiclassical
technique in terms of the 2-point Green's function on classical
trajectories; the method was first suggested in \cite{Svid}
(``$t$-representation''), and in a different form developed in
\cite{She80a,She85}.  The trajectory Green's function is introduced
via the following representation of the $2\times 2$ matrix Gor'kov
Green's function \cite{GorKop73}:
\begin{equation}
G_{\varepsilon }^{R}(\bbox{r}_{1}, \bbox{r}_{2}) = 
- {m_{F}\over 2\pi  } \;
{e^{i p_{F}|\bbox{r}_{1}-\bbox{r}_{2}|}\over
|\bbox{r}_{1}-\bbox{r}_{2}| }
\gR_{+}(\bbox{r}_{1},\bbox{r}_{2}; \varepsilon )
+ {m_{F}\over 2\pi  } \; 
{e^{-i p_{F}|\bbox{r}_{1}-\bbox{r}_{2}|}\over
|\bbox{r}_{1}-\bbox{r}_{2}| }
\gR_{-}(\bbox{r}_{1},\bbox{r}_{2}; \varepsilon )
\; , \;  p_{F}|\bbox{r}_{1}-\bbox{r}_{2}| \gg 1
\label{qza}
\end{equation}
where $m_{F}= p_{F}/v$, $p_{F}$ and $v$ being the Fermi momentum and
velocity, respectively; $\varepsilon $ in Eq.(\ref{qza}) is the energy
variable (stationary case).  For definiteness, we consider the
retarded Green's function $G^{R}$ of the Keldysh technique.  To
simplify notations, we assume a spherical Fermi surface;
generalization to an anisotropic spectrum is straightforward.

Similar to Andreev's procedure, the fast ``quantum'' oscillations on
the scale $\lambdabar_{F}$ are singled out in Eq.(\ref{qza}).
Resembling Eq.(\ref{5db}), the slowly varying quasiclassical envelopes
$\gR_{\pm}(\bbox{r}_{1},\bbox{r}_{2})$ obey first order differential
equations \cite{GorKop73,She80a}, the gradient term of which couples
only the points on straight lines which are obviously the classical
trajectories corresponding to a particle on the Fermi surface
\cite{curve}.  The trajectory is specified by its direction $\bbox{n}$
and arbitrarily chosen initial point $\bbox{R}$, so that the position
$\bbox{r}$ of a point on the trajectory $\bbox{R}, \bbox{n}$ can be
presented as $\bbox{r}= \bbox{R}+ x \bbox{n}$, $x$ has the meaning of
the coordinate on the trajectory. In the momentum space, the
trajectory $\bbox{n}$ is associated with the points in the vicinity of
the Fermi surface where the velocity vector is directed towards
$\bbox{n}$.

For the trajectory specified by $\{\bbox{n}, \bbox{R}\}$, one defines
the 2-point Green's function
$\hat{g}^{R}(x_{1},x_{2}|\bbox{n},\bbox{R})$ \cite{She80a,She85}, 
\[
\hat{g}_{\varepsilon}^{R}(x_{1},x_{2}|\bbox{n},\bbox{R})= 
\left\{
\begin{array}{rcr}
\rule[-2ex]{0ex}{0ex}
\hat{g}^{R}_{+}(\bbox{r}_{1},\bbox{r}_{2};\varepsilon)&
\;\;,\;\;&  x_{1}>x_{2} \;;\\
\hat{g}^{R}_{-}(\bbox{r}_{1},\bbox{r}_{2};\varepsilon)&
\;\;,\;\;&  x_{1}< x_{2} \; .
\end{array}
\;\; , \;\;  \bbox{r}_{1,2} = x_{1,2} \bbox{n} + \bbox{R}
\right.
\]
( In many cases we omit $\bbox{R}, \bbox{n}$ and $\varepsilon
$ for brevity and use the notation $\gR(x_{1},x_{2})$.)

As shown in \cite{She80a,She85}, the 2-point Green's function obeys 
the following equations
\begin{eqnarray}
\left( 
i v {\partial \over \partial x_{1}} 
  + 
\hat{H}_{\varepsilon,\bbox{n}}^{R}(\bbox{r}_{1})\right)
\gR_{\varepsilon}(x_{1},x_{2}|\bbox{n},\bbox{R})     & =  &
i v  \delta(x_{1} - x_{2}) 
\;\; , \;\;  \bbox{r}_{1}= \bbox{R}+ x_{1}\bbox{n}
          \label{vra1}\\   
\gR_{\varepsilon}(x_{1},x_{2}|\bbox{n},\bbox{R})
\left( 
-i v {\partial \over \partial x_{2}} 
  + 
\hat{H}_{\varepsilon,\bbox{n}}^{R}(\bbox{r}_{2})  \right)  
   &  = &
i v  \delta(x_{1} - x_{2}) 
\;\; , \;\;  \bbox{r}_{2}= \bbox{R}+ x_{2}\bbox{n}
\label{vra}
\end{eqnarray}
where the $2\times 2$ traceless \cite{traceless} matrix
$\hat{H}_{\varepsilon, \bbox{n}}^{R}$,
\[
\hat{H}_{\varepsilon, \bbox{n}}^{R}=  
 \hat{h }_{\varepsilon, \bbox{n}}^{R}
- \hat{\Sigma }_{\varepsilon, \bbox{n}}^{R}
\;\; ,  
\]
\begin{equation}
\hat{h}_{\varepsilon, \bbox{n}}^{R}=
\left(
\begin{array}{lr}
\varepsilon - \bbox{v\cdot p}_{s}&  \Delta_{\bbox{n}}\\
-\Delta^{*}_{\bbox{n}}& -\varepsilon + \bbox{v\cdot p}_{s}
\end{array}
\right) 
\;\; , \;\;  \bbox{v}= v \bbox{n} \; ,
\label{u6a}
\end{equation}
where $\Delta_{\bbox{n}}$ is the order parameter (which may dependent on
the  direction $\bbox{n}$), and 
$\bbox{p}_{s}= - {e\over c}\bbox{A}$, $\bbox{A}$ being the vector potential, 
and $\hat{\Sigma}^{R}$ is built of
the impurity self-energy  and the part of the electron-phonon
self-energy not included to the self-consistent filed $\Delta $ and .

The boundary condition to Eqs.(\ref{vra1}), and (\ref{vra}) is the
requirement that $\gR$ is zero at $|x_{1}-x_{2}| \rightarrow \infty $,
so that $\gR$ is an analytic function of $\varepsilon $ in the upper
half plane for any $x_{1,2}$ including $|x_{1}-x_{2}|= \infty $.

The advanced Green's function $\hat{g}^{A}$ is found from
Eqs.(\ref{vra1}) and Eq.(\ref{vra}) with $\hat{H}^{R}$ substituted for
$\hat{H}^{A}$,
\begin{equation}
\hat{H}^{A} = \hat{\tau}_{z} 
\left(\hat{H}^{R} \right)^{\dagger} \hat{\tau}_{z}
\label{i5a}
\end{equation}
where $\hat{\tau}_{z}$ is the Pauli matrix and the dagger denotes the
Hermitian conjugation.

Although the observables can be expressed via the quasiclassical
1-point Green's function ($x_{1}= x_{2}$), the 2-point Green's
function turns out to be a useful intermediate object.  It gives a
full physical description of the system in the approximation where the
part of the orbital degree of freedom is treated classically (no
quantum broadening in the plane $\perp \bbox{n}$), with a complete
quantum treatment of the electron-hole degree of freedom.

It is important that the construction based on the notion of smooth
classical trajectories remains valid in the presence of disorder (or
phonons), in the standard approximation when the scattering is
included on the average via the self-energy (provided $p_{F}l \gg 1$,
$l$ being the mean free path).

\subsection{ Factorization }\label{2point}

To build the Green's function on the trajectory $\bbox{n}, \bbox{R}$, one first considers solutions to the
equation
\begin{equation}
\left(iv{\partial\over{\partial x}}+\hat{H}^{R}(x)\right) \phi=0 
\label{xra}
\end{equation}
here $\phi $ is a column, $\phi = \Hcolumn{u}{v}$ and 
$\hat{H}^{R}$ stands for
$\hat{H}_{\varepsilon,\bbox{n}}^{R}(\bbox{r})$ at the
trajectory point 
$\bbox{r}= x\bbox{n} + \bbox{R}$.

Denote $\bar{\psi }$ the row built from a column $\psi $ by the
following rule:
\[
\bar{\psi } \equiv \psi^{T} \tau_{y} {1\over i}\;\; \Rightarrow  \;\;   
\overline{
\left(
\begin{array}{c}
  u     \\
   v
\end{array}
\right)}
= 
\left(
\begin{array}{ccc}
  v & ,& -u
\end{array}
\right)\;\; .
\]
Note the identities,
\[
\bar{\psi }_{a}\psi_{b}= -
\bar{\psi }_{b}\psi_{a}
\;\; , \;\;  
\bar{\psi }_{a}\psi_{a}=0 
\;\; , \;\;  
\psi_{a}\bar{\psi }_{b} -
\psi_{b}\bar{\psi }_{a}=
\left(\bar{\psi }_{b}\psi_{a} \right)\; \hat{1}
\]

By virtue of the identity 
\begin{equation}
\left(\hat{H}^{R}\right)^{T}= - \tau_{y} \hat{H}^{R} \tau_{y} \;,
\label{zra}
\end{equation}
the row $\bar{\phi }(x)$ built from a solution to Eq.(\ref{xra}),
satisfies
the conjugated equation
\begin{equation}
\bar{\phi }(x) \left(-iv{\partial\over{\partial x}}+\hat{H}^{R}(x)\right)= 0
\label{z4a}
\end{equation}

Combining Eqs.(\ref{xra}),  and (\ref{z4a}), one gets the conservation law, 
\begin{equation}
{d \over{dx}} \left(\bar{\psi_{a}} \psi_{b}\right) =0 \; ,
\label{1ra}
\end{equation}
valid for any pair of solutions $\phi_{a}(x)$ and $\phi_{b}$.

For a general complex $\varepsilon $, the Green's function is built of
the regular solutions to Eq.(\ref{xra}), {\it i.e.}  solutions
satisfying the following boundary conditions
\begin{equation}
\begin{array}{c}
\phi_{+}(x) \rightarrow 0 \;\; , \;\;  x \rightarrow + \infty   \\
\phi_{-}(x) \rightarrow 0 \;\; , \;\;  x \rightarrow - \infty 
\end{array}
\label{3ra}
\end{equation}
Denote $\phi_{\pm}^{(N)}$ the normalized solutions for which
\begin{equation}
\overline{\phi}_{-}^{(N)}(x) \; \phi_{+}^{(N)}(x) =1 \;.
\label{8ra}
\end{equation}
The normalization is possible because the l.h.s. is a
(finite) constant  as it is seen from Eq.(\ref{1ra}).

The Green's function 
can be written now as
\begin{equation}
\gR(x_{1}, x_{2})=
\left\{
\begin{array}{rcr}
\phi_{+}^{(N)}(x_{1})\; \ofi_{-}^{(N)}(x_{2})&\;\;,\;\;&  x_{1}> x_{2}  \;;\\
\phi_{-}^{(N)}(x_{1})\; \ofi_{+}^{(N)}(x_{2})  &\;\;,\;\;&  x_{1}< x_{2}     \;.
\end{array}
\right.
\label{7ra}
\end{equation}
Indeed, it satisfies Eq.(\ref{vra1}) and Eq.(\ref{vra}) at $x_{1}\neq
x_{2}$, and is regular at $|x_{1}-x_{2}|\rightarrow \infty $.  The
normalization in Eq.(\ref{8ra}) ensures that the discontinuity at
$x_{1}= x_{2}$,
\[
\gR(x+0,x)-\gR(x-0, x)= \hat{1} \; ,
\]
is what is required
by the $\delta$-function source in Eqs.(\ref{vra1}),
 and (\ref{vra}).

For clean superconductors with inelastic scattering ignored,
$\Sigma^{R(A)} \rightarrow 0$, and Eq.(\ref{xra}) is nothing but the
Andreev equation Eq.(\ref{5db}).  Note that the structure of the
equations is not changed when the disorder and inelastic scattering is
included via the self energies.  In this case, however, solutions to
Eq.(\ref{xra}) have only the meaning of the building block of the
Green's functions.

\subsection{1-point Green's function }\label{1point}

Observables can be expressed via the Green's functions with coinciding
spatial arguments, and therefore, the 1-point Green's function is the
final goal of calculations.

The 1-point Green's   functions defined as
$\gR_{\pm}(x)= \gR(x\pm 0, x)$, 
can  expressed via the normalized solutions (see Eq.(\ref{7ra}))
\begin{equation}
\gR_{+}(x) = 
\phi_{+}^{(N)}(x) \ofi_{-}^{(N)}(x)
\;\; , \;\;  
\gR_{-}(x) = 
\phi_{-}^{(N)}(x) \ofi_{+}^{(N)}(x)
\; .
\label{44a}
\end{equation}

This expression can be identically written as
\begin{equation}
\gR_{+}(x) = 
{1\over \bar{\phi_{-}}(x)\phi_{+}(x)} 
\phi_{+}(x) \ofi_{-}(x)
\;\; , \;\;  
\gR_{-}(x) = 
{1\over \bar{\phi_{-}}(x)\phi_{+}(x)} 
\phi_{-}(x) \ofi_{+}(x)
\; ,
\label{s6a}
\end{equation}
where the normalization of the wave functions $\phi_{\pm}$
is arbitrary. 

These matrices are projectors,
\begin{equation}
\gR_{\pm}\gR_{\pm}=\pm\gR_{\pm}\;\; ,\;\; 
\gR_{\pm}\gR_{\mp} = 0 \;\; , \;\;
\gR_{+} - \gR_{-} = \hat{1}
\;\; , \;\;  
{\rm Sp} \; \gR_{\pm} = \pm 1 
\label{54a}
\end{equation}

Tagging electron- and hole-like excitations in accordance with the
direction of their propagation ($\pm x$ directions) and considering
examples, e.g. the normal state, one concludes that $\gR_{+}$ can be
identified as the (quasi)electron part of the Green's function, and
$\gR_{-}$ is the (quasi)hole one (and vice versa for
$\hat{g}^{A}_{\pm}$).

Denoting
\begin{equation}
a\equiv {u_{-}\over v_{-}}\;\; , \;\;  
b \equiv  {v_{+}\over u_{+}}, 
\label{dsa}
\end{equation}
where $u_{\pm}$ and $v_{\pm}$ are the components of 
$\phi_{\pm}$,
\[
\phi_{\pm}(x)= \Hcolumn{u_{\pm}(x)}{v_{\pm}(x)}
\;\; ,
\]
Eq.(\ref{s6a}) becomes
\begin{equation}
\gR_{+} =  {1\over 1-ab}\Hcolumn{1}{b}\Hrow{1}{-a}
\;\; , \;\;  
\gR_{-} =  
{1\over 1-ab}\Hcolumn{a}{1}\Hrow{b}{-1}
\label{94a}
\end{equation}

Another elucidating form of Eq.(\ref{s6a}) is as follows
\[
\gR_{+} = \hat{O}_{a,b}
\left(
\begin{array}{lr}
1   & 0   \\
  0 & 0
\end{array}
\right)
\hat{O}_{a,b}^{-1}
\;\; , \;\;
\gR_{-} = \hat{O}_{a,b}
\left(
\begin{array}{lr}
0   & 0   \\
  0 & -1
\end{array}
\right)
\hat{O}_{a,b}^{-1}
\]
where the rotation matrix  $\hat{O}_{a,b}$
\begin{equation}
\hat{O}_{a,b}= \left(
\begin{array}{lr}
1   & a   \\
 b  & 1
\end{array}
\right)\; .
\label{f5a}
\end{equation}

As discussed in \cite{She80a,She85}, the 1-point  Green's (``$\xi-$integrated'')
function of the quasiclassical theory , $\gR$, is given by
\begin{equation}
\gR = \gR_{+} + \gR_{-} \; .
\label{04a}
\end{equation}
{\it i.e.}  
\begin{equation}
\gR =
\phi_{+}^{(N)}\ofi_{-}^{(N)}+ \phi_{-}^{(N)} \ofi_{+}^{(N)}
\;\; .
\label{w5a}
\end{equation}

In terms of $\gR$,
\begin{equation}
\gR_{\pm} = {1\over 2} \left(\gR \pm 1 \right)
\; ,
\label{a5a}
\end{equation}
and  the relations in Eq.(\ref{54a})
lead to the well-known normalization condition
\[
\left(\gR \right)^{2} = \hat{1} 
\]
and 
\[
{\rm Sp}\; \gR = 0 \;\; .
\]

Combining Eqs.(\ref{04a}) and (\ref{94a}), one gets
\begin{equation}
\gR= 
{1\over 1- ab}
\left(
\begin{array}{lr}
 1 + ab  &  -2 a   \\
 2b  & - (1 + ab)
\end{array}
\right)
\label{fsa}
\end{equation}

This parameterization of the Green's function has been recently
suggested by Schopohl and Maki \cite{SchMak95} (see, also,
\cite{Sch98}). The present derivation leads quite naturally to this
decomposition, and clearly shows the physics behind it.  Seeing that
$a$ and $b$ may be interpreted as the ``local'' amplitudes of the
Andreev reflection for electron and hole (see below) , we call them
the Andreev amplitudes.

Finally, the rotation with the matrix $\hat{O}_{a,b}$ in
Eq.(\ref{f5a}) diagonalizes $\gR$, {\it i.e.}
\[
\gR = 
\hat{O}_{a,b}
\left(
\begin{array}{lr}
1   & 0   \\
 0  & -1
\end{array}
\right)
\hat{O}_{a,b}^{-1}
\]

The advanced Green's function $\hat{g}^{A}$ and symmetry relations
between $\gR$ and $\hat{g}^{A}$ are discussed in Appendix \ref{ga}.

\subsection{Solving the equation of motion}\label{motion}

In this paper we take the approach where the main object of interest
is the two component ``wave functions'' $\phi_{\pm}$, which factorizes
the Green's function and obeys the Andreev-type equations.  A variety
of options can be chosen to find the amplitudes.  For future
references, some of them are discussed in this section.

\subsubsection{Riccati equation}\label{ric}

Instead of solving linear equations for two component $\phi =
\Hcolumn{u}{v}$, one solves the equation for the ratio $\alpha =
v/u$. It follows from Eq.(\ref{z4a}) that $\alpha (x)$ satisfies the
Riccati equation,
\begin{equation}
i {\partial\over{\partial x}} \alpha  =
2\varepsilon^{R}\alpha  + \Delta ^{*R}
+ \Delta^{R}\alpha^{2}
\label{hsa}
\end{equation}
where parameters $\varepsilon^{R}$ and $\Delta^{R}$ are found from the
identification
\[
\hat{H}^{R}(x)= 
\equiv
\left(
\begin{array}{lr}
\varepsilon^{R}&  \Delta^{R}\\
- \Delta ^{*R}   &- \varepsilon^{R}
\end{array}
\right)_{x}
\;
\]
In the context of the quasiclassical theory, 
this equation has been
first derived
by Schopohl and Maki \cite{SchMak95}.

Known  $\alpha (x)$, one finds the 2-component function $\phi (x)$,
\begin{equation}
\phi(x)= 
const\;
\left(
\begin{array}{c}
    1   \\
   \alpha(x)
\end{array}
\right)
 \exp\left(i 
\int\limits_{x_{0}}^{x} dx'\left(\varepsilon^{R}
 + \Delta^{R}\alpha\right)_{x'}\right)
\label{jsa}
\end{equation}

To find $\alpha_{\pm}(x)$ {\it i.e.}  the solutions to Eq.(\ref{hsa})
corresponding to $\phi_{\pm}\;$, the Riccati equation must be
supplemented with the boundary condition which leads to the correct
asymptotics  Eq.(\ref{3ra}).

In many cases of interest such e.g. an SNS-structure or isolated
Abrikosov's vortex, the superconductor is homogeneous at $x
\rightarrow \pm \infty $.  If so, solutions to Eq.(\ref{xra}) are
plane waves in the asymptotic region:
\[
\phi (x) \rightarrow conts \, \Hcolumn{\Delta^{R}}
{\pm \xi^{R} -\varepsilon^{R} } e^{\pm i\xi^{R}x}
\]
$\xi^{R}= \sqrt{(\varepsilon^{R})^{2}- \Delta^{R} \Delta^{*R} }\;,\;
\Im \xi^{R}>0$.  Selecting the waves decaying in the corresponding
region, one comes to the boundary conditions as follows:
\begin{equation}
 \phi_{\pm}:  \hspace{2ex}    
\alpha_{\pm}|_{x = \pm\infty } = 
{\pm \xi^{R}- \varepsilon^{R}\over \Delta^{R}}|_{x=\pm \infty }
\label{q5a}
\end{equation}
An equivalent condition was suggested in \cite{SchMak95,Sch98} from ``the
requirement of the stability of the numerical integration procedure''.
In the present paper, the boundary condition is deduced, ultimately, from the physical
condition that the 2-point Green's function is a regular function
decaying at large distance from the source.

The 1-point Green's functions $\gR_{\pm}$ and $ \gR$ are
found now from Eq.(\ref{94a}) and Eq.(\ref{fsa}) with the
understanding that
\[
b(x) = \alpha_{+}(x)\;\; , \;\;  
a(x) = 1/ \alpha_{-}(x)
\]
where $\alpha_{\pm}(x)$ are the solutions to Eq.(\ref{hsa}) with the
boundary conditions in Eq.(\ref{q5a}).

\subsubsection{Periodic potential}\label{per}

In many situations of interest such us vortex lattice, N-S or S-S
superlattice, or multiple reflections (see below) the potentials are
periodic functions of the trajectory coordinate.  In this case, the
Green's functions may be found by the following method.

A formal solution to Eq.(\ref{xra}),
$\phi(x) = \hat{U}(x,x_{0})\phi(x_{0})$,
can be expressed via the evolution  matrix 
\[
\hat{U}(x,x_{0})=
T_{x}
e^{-i \int\limits_{x_{0}}^{x_{1}}dx'\;\hat{H}^{R}(x') }
\]
where  $T_{x}$ orders the matrices $\hat{H}^{R}(x)$ in the descending
$x-$order from the left to the right.

Denote $\hat{U}_{L}(x)\equiv \hat{U}(x+L,x) $ the evolution matrix
corresponding to the translation by the period of the structure $L$.
As proven in Section \ref{append.period}, the 1-point Green's function
can be found as
\begin{equation}
\gR(x)= \Format{1.2}{\hat{U}_{L}(x)}\; .
\label{u5a}
\end{equation}
Here 
$\Format{0.7}{\ldots}$ 
stands for the ``formating'' operation: 
\begin{equation}
\Format{1.2}{\hat{Q}} = 
{1\over q_{R} }
\left(\hat{Q} - \Big({1\over 2}{\rm Sp}\, \hat{Q} \Big) \hat{1}  \right)
\;\; , \;\;  
q_{R} = \sqrt{\left(\hat{Q} - 
\Big({1\over 2}{\rm Sp}\, \hat{Q} \Big) \hat{1}  \right)^{2}}
\; ,
\label{y7a}
\end{equation}
which returns a normalized traceless matrix \cite{format} (similar
combination of matrices has been introduced in \cite{Nag98}).  The
branch of the square root in $q_{R}$ must be chosen to satisfy
$\Re\left(\Format{0.7}{\hat{Q}} \right)_{11}>0$.  
Except for the
choice of the branch, Eqs.(\ref{u5a}), and (\ref{y7a}) are same for
$\hat{g}^{A}$.  Construction of the evolution matrix $\hat{U}_{L}(x)$
in the Riccati equation technique is described in Section
\ref{append.period}.

\section{Knot matching conditions}\label{knot}

In the quasiclassical picture, particles move on trajectories,
usually, straight lines characterized by the direction of velocity
$\bbox{n}$ (and the initial position $\bbox{R}$). At any point in real
space, infinite number of trajectories with different $\bbox{n}$ cross
each other.  Since there is no transitions between the intersecting
trajectories, the crossings do not lead to any physical effect. At
some points, called here knots, the quasiclassical condition is
violated. At a knot, the particle may leave its original trajectory and continue
its motion along a trajectory in another direction. In the simplest
example of a specular interface Fig. \ref{cross}, two trajectories
1-1' and 2-2' are mixed.  In a general case, the knot is a region
where transitions between $N$ in- and $N$ out-trajectories are
allowed.  The in- trajectories (or channels) are those which have the
direction of the Fermi momentum towards the knot; the momentum
direction is from the knot in the out-channels (see Fig. \ref{3-Fig})
\cite{irred}.  The in- and out-trajectories are somehow numbered, $l=
1,\ldots , N$.  We mark by $'$ the out-going channels so that $k'$
stands for the k-th outgoing channels.

Since
the knot is
point-like
on the quasiclassical scale $\sim v_{F}/ \Delta $, one can talk about
the knot value of the trajectory ``wave function''. Denote $\psi_{i}$
the 2-component wave function on the $i$-th in-coming trajectory, $i=
1,\ldots ,N$ at the point where it enters the knot, and analogously
$\psi_{k'}$ i the knot value on the $k-$th outgoing trajectory.

The outcome of events happening inside the knot can be generally
described by the scattering S-matrix.  For any specified case, it can
be found by solving the Schr\"odinger equation for the electron with the
Fermi energy.  Here, it is considered as a phenomenological input.

The suggested matching condition reads
\begin{equation}
\psi_{k'} = \sum\limits_{i=1}^{N} S_{k'i}\psi_{i} \; ,
 \label{wsa}
\end{equation}
where $S_{k'i}$ are the elements of the unitary scattering matrix.  In
the spirit of the quasiclassical theory, $S_{k'i}$ is the normal metal
property taken at the Fermi surface; it is an electron-hole scalar.
This relation generalizes the matching conditions of Ref.\cite{She80b}
to the many channels case.

Taking advantage of unitarity, $S^{-1}= S^{\dagger}$, the inverse of
Eq.(\ref{wsa}) reads
\begin{equation}
\psi_{i} = \sum\limits_{k'=1}^{N} S_{ik'}^{\dagger}\psi_{k'}
\label{zsa}
\end{equation}

Seeing that the conjugated wave function $\ops$ always belongs to the
second argument of the Green's function $G(1,2)\sim \langle \psi (1)
\psi^{*}(2) \rangle $, it must obey the matching conditions for
$\psi^{*} $ {\it i.e.}
\begin{equation}
\ops_{k'} = \sum\limits_{i=1}^{N} S^{*}_{k'i}\ops_{i}
\label{75a}
\end{equation}

Eq.(\ref{xra}) together with the matching conditions in
Eq.(\ref{wsa}), allows one to find the 2-component amplitudes on the
tree-like trajectory, and, therefore, the Green's functions.  We
remark also that the relation in Eq.(\ref{wsa}) can be used as the
boundary condition to the Andreev equation \ref{5db}.

\section{ Andreev reflection on the knot}\label{andreev}

In this section, we consider the quantum problem of scattering of
ballistic excitations off the knot or, in other words, the problem of
many-channel combined, Andreev and usual, reflection/transmission. The problem is
formulated as follows.  On each of the trajectories connected by the
knot, $i,k'= 1,2,\ldots N$, the 
order parameter $\Delta(x) $ and, hence,
the matrix 
$\hat{h}(x)$
in Eq.(\ref{u6a}) is supposed to be known.  Since in the ballistic
case $\Sigma =0$, the wave function on each of the trajectories
satisfies the equation
\begin{equation}
\left(iv{\partial\over{\partial x}}+\hat{h}(x)\right) \psi =0
\;\; ,
\label{v6a}
\end{equation}
here $x$ is the coordinate along the corresponding trajectory; this
equation differs only in notations from the Andreev equation
Eq.(\ref{5db}).  The scattering of the (quasi)particles off the knot
is due multiple sequential processes of (i) inter-trajectory
transitions described by Eq.(\ref{wsa}), which do not affect the the
electron-hole degrees of freedom, followed by (ii) intra-trajectory
Andreev reflections i.e.  rotations in the electron-hole space.  The
goal is to express the amplitudes of the multiple processes via the
amplitudes of the elementary events.

On each of the paths, we chose the origin $x=0$ at the knot. Then, the
coordinate $x$ belongs to the region $- \infty <x<0$ on the in-coming
and to the region $0< x< \infty $ on the outgoing trajectories.

First, we consider  the plane wave
asymptotics at $|x| \rightarrow \infty $
where
$\hat{h}= const(x)$. 
The electron-like 
(hole-like) solution is 
$\Psi_{e}(x)=\psi_{e} e^{i \xi x/v}$ 
($\Psi_{h}(x)=\psi_{h}e^{-i \xi x/v}$),
where $\psi_{e}$ ($\psi_{h}$), 
$\hat{h}\psi_{e} = +\xi\psi_{e}$ 
($\hat{h}\psi_{h} = -\xi\psi_{h}$)
is the eigenfunction of the matrix $\hat{h}$. 
The eigenvalues $\pm\xi $ are found from $ \xi^{2} \hat{1}=
\hat{h}^{2}$.  We supply the energy with an infinitesimal {\em
positive} imaginary part, $\varepsilon \rightarrow \varepsilon + i
\delta $, and impose condition $\Im \xi >0$ to specify the branch of
$\sqrt{\xi^{2}}$.

The basis for the electron-hole classification is the quasiparticle
current
\begin{equation}
j_{qp} = \psi^{\dagger}\hat{\tau}_{z}\psi = |u|^{2} - |v|^{2} \;  
\label{t6a}
\end{equation}
which is a constant of motion, ${d \over{dx}}j_{qp}(x) = 0$, due to
the symmetry $\hat{h}^{\dagger}= \hat{\tau}_{z}\hat{h}\hat{\tau}_{z}$.
The electron-like quasiparticle is identified by $j_{qp}>0$.  It moves
in the direction of increasing $x$ in accordance with the sign of the
probability current.  For the hole-like excitation $j_{qp}<0$, and it
moves towards $x=- \infty $. Note that the solution $\Psi^{(e,h)}$ are
chosen in the way that both electron and holes decay in the direction
of propagation.

Below, $\psi^{(e,h)}$ denotes the
eigenfunctions 
normalized to the unit flux:
\[
\psi^{(e)\dagger}\hat{\tau}_{z}\psi^{(e)} = 1
\;\; , \;\;  
\psi^{(h)\dagger}\hat{\tau}_{z}\psi^{(h)} = -1 
\;\; , \;\;  
\xi^{2}>0 \;\;.
\]
(The l.h.s. is identically zero
in the gap region when $\xi^{2}< 0$ and 
 propagating states are absent.)

Generally, $\hat{h}$ is $x$-dependent and the solutions are the plane
waves only asymptotically. However, the electron-hole classification
is unique due to the current conservation in Eq.(\ref{t6a}).  One has
for the electron- , $\Psi^{(e)}(x)$, and hole-like, $\Psi^{(h)}(x)$,
solutions on out-going (in-coming) trajectories
\begin{equation}
\Psi^{(e)}(x) = \left\{
\begin{array}{rcr}
\rule[-3ex]{0ex}{0ex}
\psi^{(e)}\;e^{i \xi x/v}&\;\;,\;\;& 
 x \rightarrow  \infty \; (\text{or } - \infty ) \;;\\
\displaystyle
{1\over  \beta^{(e)}} 
\Hcolumn{1}{\alpha^{(e)}} 
 &\;\;,\;\;&  x=0\;.
\end{array}
\right.
\label{y6a}
\end{equation}
\begin{equation}
\Psi^{(h)}(x) = \left\{
\begin{array}{rcr}
\rule[-3ex]{0ex}{0ex}
\psi^{(h)}\;e^{-i \xi x/v}&\;\;,\;\;&  
x \rightarrow  \infty \; (\text{or } - \infty ) \;;\\
\displaystyle
{1\over  \beta^{(h)}} 
\Hcolumn{\alpha^{(h)}}{1} 
 &\;\;,\;\;&  x=0\;.
\end{array}
\right.
\label{z6a}
\end{equation}
where the parameters $\alpha^{(e,h)}$ and $\beta^{(e,h)}$ are found
solving Eq.(\ref{v6a}) in the region $0< x < \infty $ (or $- \infty <
x < 0$). 

If considered as a function of $x$, $\alpha^{(e)}(x)$ and $1/
\alpha^{(h)}$ can be found by solving Eq.(\ref{hsa}). We see that
indeed the parameters $a(x)$ and $b(x)$ of the Riccati equation technique
have the meaning of the instantaneous (local) amplitudes of Andreev
reflection, and, therefore, one may call them the Andreev amplitudes.

It generally follows  from the current conservation Eq.(\ref{t6a})
that  
\[
\alpha^{(h)} = \left(\alpha^{(e)} \right)^{*}
\;\; , \;\;  
|\alpha^{(e)}|^{2} + |\beta^{(e)}|^{2} =
|\alpha^{(h)}|^{2} + |\beta^{(h)}|^{2} = 1
\; .
\]
for an open channel, $ \xi^{2}>0$.
Seeing that $|\beta^{(e)}|^{2}= |\beta^{(h)}|^{2}$, one can enforce
\[
\beta^{(e)}= \beta^{(h)} \; 
\]
choosing the
overall phase factor in $\psi^{(e,h)}$.

The physical meaning of the parameters is clear from Eqs.(\ref{y6a}),
and (\ref{z6a}): On the outgoing trajectories ($0< x< \infty $),
$\alpha^{(e)}$ is the amplitude of the Andreev reflection of the
(bare) electron injected at $x=0$, and $\beta^{(e)}$ is the
corresponding transmission amplitude; $\alpha^{(h)}/\beta^{(h)}$ and
$1/\beta^{(h)}$ are $u-, v-$ components of the quasi-hole having come
from $x= \infty $.  On the in-coming paths, the above is true after
the substitution ``electron'' $\leftrightarrow$ ``hole''.

Moving towards the knot, quasi-electrons on the in-coming and
quasi-holes on the out-going trajectories comprise the in-coming
states of the scattering problem; the out-going states are electrons
on the out-going and holes excitations on the in-coming trajectories.

Let the incoming particle be the quasi-electron approaching the knot
along the $l$-th in-trajectory.  The source particle generates waves
in {\em all} out-going channels.  The wave functions of the system
$\Psi ^{(l)}$ reads
\begin{equation}
\Psi ^{(l)} = 
\Psi^{(e)}_{l} 
+ B_{l}^{(l)}\Psi^{(h)}_{l}
+ \sum\limits_{k\neq l} B_{k}^{(l)}\Psi^{(h)}_{k}
+ \sum\limits_{ k'}
A_{k'}^{(l)}\Psi^{(e)}_{k'}
\label{36a}
\end{equation}
where $\Psi^{(e)}_{k'}$ and $\Psi^{(h)}_{k}$ stands for the trajectory
wave functions defined by Eqs.(\ref{y6a}), and (\ref{z6a}), $k$ or
$k'$ being the label of the trajectory.  The yet unknown amplitudes of
the outgoing particles, $A_{k'}^{(l)}$ and $B_{k}^{(l)}$, are to be
found from the matching conditions in Eq.(\ref{wsa}).

The calculations are most easily done using Eqs.(\ref{d6a}), and
(\ref{f6a}).  It follows by comparing Eq.(\ref{d6a}) with
Eqs.(\ref{y6a}), (\ref{z6a}), that one may put $\nu_{m\neq l}=
\alpha^{(h)}_{m}$ and $\mu_{k'}= \alpha^{(e)}_{k'}$.  From
Eq.(\ref{d6a}) and Eq.(\ref{b6a}), one sees that the wave functions on
the source trajectory at $x_{l}=0$ must be proportional to ${1\choose
\alpha_{0l}^{(e)}}$ , where
\begin{equation}
\alpha_{0l}^{(e)} = 
\langle l|S^{\dagger} \hat{\alpha}^{(e)} \hat{\bbox{S}}_{l}|l \rangle
.
\label{96a}
\end{equation}
Here and below, $\hat{\bbox{S}}_{l}$, is the full S-matrix taking into
account multiple events of the Andreev reflection. From Eq.(\ref{c6a})
\begin{equation}
\hat{\bbox{S}}_{l} =
\left(\hat{S}^{\dagger} - (\hat{\alpha}^{(h)})^{(l)}
\hat{S}^{\dagger}\hat{\alpha}^{(e)}  \right)^{-1}\;\; ;
\label{56a}
\end{equation}
where $\hat{\alpha}^{(e,h)}$ is the diagonal matrix with the elements
$\left(\hat{\alpha}^{(e,h)} \right)kk = \alpha^{(e,h)}_{k}$ and
superscripts $^{(l)}$ means the $ll$-element must be put to zero.  By
$\langle l|Z| m\rangle $, $l,m = 1,\ldots,N$ we denote the matrix
element $Z_{lm}$.

The parameter $\alpha_{0l}^{(e)}$ in Eq.(\ref{96a}) has the meaning of
the amplitude of the Andreev backscattering of a bare electron by the
knot as a whole.  From the condition that the wave function has the
$u-v$ structure at $x_{l}=0$ like ${1\choose \alpha_{0l}^{(e)}}$, one
finds $B_{l}$, {\it i.e.}  the amplitude of the Andreev reflection of
the incident electron excitation.  After some algebra
\begin{equation}
B_{l}^{(l)} = 
{
\alpha_{0l}^{(e)} - \alpha_{l}^{(e)}\over 
1 - \alpha_{0l}^{(e)} \alpha_{l}^{(h)}
}\; .
\label{06a}
\end{equation}
Here, the denominator can be understood as due to multiple Andreev
reflections \cite{real}.

The wave function $\Psi^{(e)}_{l}+ B_{l}^{(l)}\Psi^{(h)}_{l}$ 
at $x_{l}=0$ equals now to 
$C{ 1 \choose  \alpha_{0l}^{(e)}  }$ where 
$C = \beta_{l}^{(e)*}\left(1 - \alpha_{0l}^{(e)} \alpha_{l}^{(h)}
\right)^{-1} $.
Looking at Eq.(\ref{d6a}), one find the rest of the scattering amplitudes: 
\begin{equation}
A_{k'}^{(l)}= 
{1 
\over 1 - \alpha_{0l}^{(e)} \alpha_{l}^{(h)}}
\langle k'|\hat{\beta}^{(e)} 
\hat{\bbox{S}}_{l} \hat{\beta}^{(e)*}|l\rangle 
\;\; , \;\;  
B_{k}^{(l)} = 
{1 
\over 1 - \alpha_{0l}^{(e)} \alpha_{l}^{(h)}}
\langle k|
\hat{\beta}^{(h)}
S^{\dagger} \hat{\alpha}^{(e)} \hat{\bbox{S}}_{l}
\hat{\beta}^{(e)*}
|l \rangle
\;\; .
\label{46a}
\end{equation}

Similarly, one derives the scattering amplitudes for the quasi-hole coming to
the knot on the $n'$-trajectory. Analogously to Eq.(\ref{36a}), the
wave function,
\[
\Psi ^{(n')} =
\Psi^{(h)}_{n'}
+ B_{n'}^{(n')}\Psi^{(e)}_{n'}
+ \sum\limits_{k'\neq n'} B_{k'}^{(n')}\Psi^{(e)}_{k'}
+ \sum\limits_{ k}
A_{k}^{(n')}\Psi^{(h)}_{k}\; ,
\]
contains the scattering amplitudes which are found from the matching
conditions.  The corresponding expressions can be obtained by the
substitutions: $(e) \leftrightarrow (h) $, $l \rightarrow n'$, and $S
\leftrightarrow S^{\dagger}$, and $\bbox{S}_{l}\rightarrow
\bbox{S}_{n'}^{\dagger}$,
\[
\hat{\bbox{S}}_{n'}^{\dagger}= 
\left(\hat{S} - (\hat{\alpha}^{(e)})^{(n')} \hat{S}
\hat{\alpha}^{(h)}\right)^{-1}\; .
\]

For the hole incident on the
$n'$-trajectory, 
the amplitudes of the Andreev reflection, $B_{n'}^{(n')}$, scattering
to the hole state on the $k$-th trajectory, $A_{k}^{(n')}$, and
scattering to the electron state on the $k'$-th trajectory,
$B_{k'}^{(n')}$, read respectively
\begin{mathletters} 
\begin{eqnarray}
B_{n'}^{(n')}& =& 
{
\alpha_{0n'}^{(h)} - \alpha_{n'}^{(h)}\over 
1 - \alpha_{0n'}^{(h)} \alpha_{n'}^{(e)}
}\; , 
\\
 A_{k}^{(n')}&=&
 {1
\over 1 - \alpha_{0l}^{(h)} \alpha_{l}^{(e)}}
\langle k|\hat{\beta}^{(h)} 
\hat{\bbox{S}}_{n'}^{\dagger}\hat{\beta}^{(h)*}|n'\rangle
\\
B_{k'}^{(n')}& = &
{1
\over 1 - \alpha_{0l}^{(h)} \alpha_{l}^{(e)}}
\langle k'|
\hat{\beta}^{(e)}
S \hat{\alpha}^{(h)} \hat{\bbox{S}}_{n'}^{\dagger}
\hat{\beta}^{(h)*}
|n' \rangle
\label{i7a}
\end{eqnarray}
\end{mathletters}

The presented formulae give the amplitude of scattering
from a propagating channel to another propagating channel.
The scattering of the excitations is a result of multiple sequential
events of two types: (i) on the knot inter-trajectory transitions
described by the $S$-matrix in Eq.(\ref{wsa}), and (ii)
intra-trajectory processes of the Andreev reflection/transmission with
the amplitudes
$\alpha^{(e,h)}$/$\beta^{(e,h)}$. Expanding the effective $S$-matrix $\bbox{S}_{l}$ in
Eq.(\ref{56a}), $\bbox{S}_{l} = 
\hat{S} + 
\hat{S}(\hat{\alpha}^{(h)})^{(l)}\hat{S}^{\dagger}\hat{\alpha}^{(e)}\hat{S}
+
\hat{S}
(\hat{\alpha}^{(h)})^{(l)}\hat{S}^{\dagger}\hat{\alpha}^{(e)}\hat{S}
(\hat{\alpha}^{(h)})^{(l)}\hat{S}^{\dagger}\hat{\alpha}^{(e)}\hat{S}
+ \ldots
$  
one sees that the  full amplitude of the scattering event $n'
\leftarrow m$  is the superposition of all different paths connecting
the initial and final states with electron $\leftrightarrow$ hole
transformation on each  step.

The theory gives exact amplitudes of the multiple scattering expressed
via the amplitudes of the elementary processes: the normal metal
$S$-matrix and the intra-trajectory Andreev amplitudes.
In the simplest case, when $N=2$, and $\Delta =0$ on two out of the 4
trajectories, the above formula reproduce results
of the theory of Andreev reflection in the NIS structure \cite{She80a}.

\subsection{Bound states}\label{bound}

Bound states are physical solutions existing in the absence of a
source.  The physical solutions are those when the matching conditions
on the knot are simultaneously satisfied with the requirement that the
wave functions decay far away from the knot.  The electron and holes
states defined earlier (with $\Im \xi >0$) have the property that they
decay in the direction of their propagation. Therefore, the wave
function of a bound state $\Psi_{\text{bound}}$ has the form
\[
\Psi_{\text{bound}} =
 \sum\limits_{k} B_{k}\Psi^{(h)}_{k}
+ \sum\limits_{ k'}
A_{k'}\Psi^{(e)}_{k'}
\]
where the coefficients $A$'s and $B$'s are found from the matching
conditions. Again, looking at Eqs.(\ref{y6a}), and (\ref{z6a}) one sees
that in Eq.(\ref{85a}), $\mu_{k'}$ may be identified with
$\alpha^{(e)}_{k'}$, and $\nu_{i}$ with $\alpha^{(h)}_{i}$. Then,
Eq.(\ref{05a}),
\begin{equation}
{\cal D}(\{\alpha^{(h)}\},\{\alpha^{(e)}\})\equiv
\det
\left|
\left|
1 -
\hat{S}\hat{\alpha}^{(h)} \hat{S}^{\dagger}\hat{\alpha}^{(e)}
\right|
\right|
 = 0 \; ,
\label{m7a}
\end{equation}
gives the condition for  the wave functions to be matched 
on the knot. 
The Andreev amplitudes $\alpha^{(e)}$ and $\alpha^{(h)}$ are functions
of energy $\varepsilon $,
and the bound states exist at the energies  where Eq.(\ref{m7a}) is
satisfied.

\subsection{Example:  Rough surface, anisotropic superconductor}\label{rough}

The rough surface reflects waves in many direction.  As the simplest
model, we assume that the surface reflection couples together only 2
in-coming directions ''1'' and ``2'' to two outgoing
``1$^{\prime}$'' and ``2$^{\prime}$''. 
The model corresponds to a $N=2$ knot.  In what follows we calculate
the amplitude of Andreev reflection by the knot and consider the bound
levels.

The unitary $2\times 2$ scattering matrix of the knot may be taken in
the form
\[
\hat{S} = 
\left(
\begin{array}{lr}
r_{1}& r_{2}\\
 -r_{2}^{*}& r_{1}^{*}
\end{array}
\right)
\]
provided $R_{1}+R_{2}=1$, where $R_{1}= |r_{1}|^{2}$ 
($R_{2}= |r_{2}|^{2}$) is the probability
of reflection $1 \rightarrow  1'$ ($2 \rightarrow 1'$).

Given the profile of the order parameter, one can find the wave
functions, and the Andreev amplitudes $\alpha^{(e,h)}$ and
$\beta^{(e,h)}$.  Here, the matrices 
\[
\hat{\alpha}^{(h)} =
\left(
\begin{array}{lr}
 \alpha^{(h)}_{1}& 0    \\
 0  & \alpha^{(h)}_{2}
\end{array}
\right)
\;\; , \;\;  
\hat{\alpha}^{(e)}=
\left(
\begin{array}{lr}
\alpha^{(e)}_{1'}& 0    \\
 0  & \alpha^{(e)}_{2'}
\end{array}
\right)\; ,
\]
are taken as input, each of the $\alpha $'s is a functions of energy.

The energies of  bound states are found from
 Eq.(\ref{m7a}), which takes the following form
\begin{equation}
{\cal D}(\varepsilon )\equiv 
R_{1}
(1 - \alpha_{1}^{(h)} \alpha_{1'}^{(e)})
(1 - \alpha_{2}^{(h)} \alpha_{2'}^{(e)})
+ R_{2}
(1 - \alpha_{1}^{(h)} \alpha_{2'}^{(e)})
(1 - \alpha_{2}^{(h)} \alpha_{1'}^{(e)})
=0
\; .
\label{v7a}
\end{equation}
The  bound states exist only in the
gap region 
at the energy interval
where 
$|\alpha_{1,2}^{(h)}|= |\alpha_{1',2'}^{(e)}|=1$. 

Essential physics
can be grasped by the simplest model where the order parameter
$\Delta_{n}$ is a constant at each of the trajectories: 
$\Delta_{n}=
 \Delta e^{i \varphi_{n}}$. Then,\[
\alpha_{1',2'}^{(e)}= e^{{i\over 2}\psi} e^{- i \varphi_{1',2'}}
\;\; , \;\;  
\alpha_{1,2}^{(h)} = e^{{i\over 2}\psi} e^{i \varphi_{1,2}}
\]
where $\psi_{\varepsilon}$ is a function of energy, $e^{i
\psi_{\varepsilon }}= (\varepsilon - i\sqrt{|\Delta |^{2} -
\varepsilon^{2}})/ (\varepsilon + i\sqrt{|\Delta |^{2} -
\varepsilon^{2}} ) $.

 Eq.(\ref{v7a}) is conveniently  transformed to the form, 
\begin{equation}
\cos ( \psi_{\varepsilon} + \varphi_{11'} + \varphi_{22'}) =
R_{1} \cos \left({\varphi_{12} - \varphi_{1'2'}\over 2} \right)
+
R_{2} \cos \left({\varphi_{12} + \varphi_{1'2'}\over 2} \right)
\; ,
\label{w7a}
\end{equation}
$\varphi_{ab}\equiv \varphi_{a}- \varphi_{b}$.

One sees that the existence and position   of the bound
state is sensitive to the surface roughness only
if either the incoming or outgoing channels are not
equivalent {\it i.e.}  
$\varphi_{12}= \varphi_{1}- \varphi_{2} \neq 0$, or
$\varphi_{1'2'}= \varphi_{1'}- \varphi_{2'} \neq 0$. 
In other words, mixing of identical channel does not 
affects the levels.

Consider now the possibility, which may exist in the case of a d-wave
superconductor, that the order parameter changes its sign on the $1
\rightarrow 1'$ and $2 \rightarrow 2'$ trajectories.  A smooth surface
mixes only trajectories with close transverse momenta; then the
trajectories are almost equivalent and their coupling does not shift
the levels.  On the contrary, a backward-like scattering splits the
degenerate levels: In the model under consideration, the backward-like
scattering corresponds to the phase factors $\varphi_{1}=
\varphi_{2'}= \pi $ and $\varphi_{2}= \varphi_{1'}= 0 $.  Then, from
Eq.(\ref{w7a}) $\cos \psi_{\varepsilon}= R_{2}-R_{1}$. The bound state
energies are
\begin{equation}
\varepsilon_{\text{bound}} = \pm \sqrt{R_{2}}\,\Delta \; . 
\label{x7a}
\end{equation}
One concludes that the presence of substantial spectral weight at low
energies is not likely if scattering in the backward directions is
present: $\sim$10\% 
probability the scattering moves the levels from zero energy to $\sim
0.3 \Delta $, of the order of the gap.

The amplitudes of scattering of excitations can be found from
Eqs.(\ref{06a}), (\ref{46a}) and (\ref{i7a}) As an example, the
amplitude of the Andreev reflection of the electron-like excitation
incident on the trajectory ``1'', $B_{1}^{(1)}$, reads 
\[
B_{1}^{(1)} =
{
R_{1}\tilde{\alpha}_{1'}^{(e)} - \tilde{\alpha}_{1}^{(e)}\over
1 - R_{1}\tilde{\alpha}_{1'}^{(e)} \tilde{\alpha}_{1}^{(h)}
}\; .
\]
where the following notations are used
\begin{eqnarray}
\tilde{\alpha}^{(h)}_{1}& =  & {\alpha^{(h)}_{1}- \alpha^{(h)}_{2}
\over 1 - \alpha^{(e)}_{2'} \alpha^{(h)}_{1}}   \; ,  
\nonumber\\   
\tilde{\alpha}^{(e)}_{1'}& =  & {\alpha^{(e)}_{1'}- \alpha^{(e)}_{2'}
\over 1 - \alpha^{(e)}_{1'} \alpha^{(h)}_{2}} 
\; ,  
\nonumber\\   
\tilde{\alpha}^{(e)}_{1}& =  & {\alpha^{(e)}_{1}- \alpha^{(e)}_{2'}
\over 1 - \alpha^{(e)}_{1} \alpha^{(h)}_{2}} 
\; .
 \nonumber 
\end{eqnarray}
The shortest way to derive this result is
to apply the rotation  transforming  $\alpha^{(h)}_{2}$ and
$\alpha^{(e)}_{2'}$ to zero as explained in Section \ref{det}.

\section{Matching Green's functions}\label{match}

As has been discussed in Section \ref{2point} and \ref{1point}, the
Green's functions can be built from the regular solutions to the Andreev
equation Eq.(\ref{xra}).  When the trajectory coordinate $x$ extends
from $- \infty $ to $\infty $, the regularity requirement leads to
the boundary conditions in Eq.(\ref{3ra}).  In the case of a
trajectory ending in or originating from a knot the boundary
conditions must be reformulated.

First consider  an isolated knot  mixing
semi-infinite trajectories
(with no more knots on them). 
With the origin chosen at the knot, the trajectory coordinate $x_{n}$
extends from $-\infty $ to $0$ on the $n$-th incoming trajectory, and
$0< x_{k'} < \infty $ on the $k'$-outgoing one. 
As before, the requirement,
\begin{equation}
\phi_{-, m}(-\infty ) = 0
\;\; , \;\;  
\phi_{+, k'}(\infty ) = 0 
\; \; , m,k=1,\ldots,N \; ,
\label{z7a}
\end{equation}
uniquely (up to a normalization factor) defines the solutions $\phi_{-,n}(x_{n})$ and
$\phi_{+,k'}(x_{k'})$. 
Denote the knot values of the regular solutions as  
\begin{equation}
\phi_{-,m}(x_{m}=0) = {\displaystyle a_{m}\choose 1}
\;\; , \;\;  
\phi_{+,k'}(x_{k'}=0) = {\displaystyle 1 \choose b_{k'}}
\; \; , m,k=1,\ldots,N \; .
\label{27a}
\end{equation}
For convenience, the normalization is chosen so that one of the
components equals to 1 at the knot;  
the parameters $a_{m}$ or $b_{k'}$ are ``bulk'' properties
independent on the knot.

The problem in hand is to find  the knot values
\[
\phi_{+,l}(x_{l}=0) \equiv {\displaystyle 1 \choose b_{l}} 
\;\; , \;\;   
\phi_{-,n'}(x_{n'}=0) \equiv {\displaystyle a_{n'}\choose 1}, 
\;\; , \;\;  l,n=1,\ldots,N
\]
which give 
 the boundary condition 
 to Eq.(\ref{xra})
needed to evaluate
$\phi_{+,l}(x_{l}<0)$ and $\phi_{-,n'}(x_{n'}>0)$.

To find $\phi_{+,l}(0)$, one notes that by virtue of the matching
conditions in Eq.(\ref{wsa}) and Eq.(\ref{zsa}), a finite
$\phi_{+,l}(0)$ generates waves in all other channels, outgoing and
incoming.  In a regular solution, all the secondary waves must decay
while propagating from the knot. This condition fixes the $u-v$
structure of the secondary waves: in each of the channel, the incoming
$m\neq l$ and any outgoing one $k'$, the generated 2-component wave
functions (at $x=0$) must be proportional to that in Eq.(\ref{27a}).
As proven in Section \ref{formal}, the matching condition allows one
to find the $u-v$ structure in one of the channels provided, as is the
case here, it is known for all other channels.

Changing notions in formulae in Section \ref{formal}
($\mu_{k'}\rightarrow b_{k'}\,$, $\nu_{i\neq l}\rightarrow a_{i}\,$,
$\nu_{l^{-1}}=b_{l}$), one gets from Eq.(\ref{b6a})
\begin{equation}
b_{l} = 
\langle l|
\hat{S}^{\dagger}\hat{b} \hat{\bbox{S}_{l}}
|l \rangle
\; 
\label{47a}
\end{equation}
where 
\[
\hat{\bbox{S}}_{l} =
\left(\hat{S}^{\dagger} - \hat{a}^{(l)}
\hat{S}^{\dagger}\hat{b}  \right)^{-1}\;\; ,
\]
$\hat{a}= \text{diag}(a_{1},a_{2},\ldots)$ and $\hat{b}=
\text{diag}(b_{1'},b_{2'},\ldots)$; the superscript $^{(l)}$ has
the meaning that the $l$-th element on the diagonal must be put to
zero; and $\langle l|(\ldots)|l \rangle \equiv (\ldots)_{ll}$.

Repeating the arguments, one finds the boundary value $a_{n'}$.
Changing notations in Eq.(\ref{a6a}) ($\mu_{n'}^{-1}\rightarrow
a_{n'}$), one gets
\begin{equation}
a_{n'}=
\langle n'|\hat{S} \hat{a }\hat{\bbox{S}}_{n'}^{\dagger}|n'\rangle
\; .
\label{67a}
\end{equation}
where
\[
\hat{\bbox{S}}_{n'}^{\dagger}= \left(\hat{S} - \hat{b}^{(n')} \hat{S}
\hat{a}\right)^{-1}\; ,
\]

From the derivation in Section \ref{formal}, it is clear that both
Eq.(\ref{47a}) and \ref{67a} are just different forms of
Eq.(\ref{05a}), 
which
 reads in 
the present
notations 
\begin{equation}
{\cal D}(\{a\},\{b\})\equiv
\det
\left|
\left|
1 -
\hat{S}\hat{a} \hat{S}^{\dagger}\hat{b}
\right|
\right|
 = 0 \; .
\label{87a}
\end{equation}
This equation should be understood in the following sense:

Suppose one seeks for the boundary value of $b_{l}$ for the $l$-the
in-channel.  Then, one formally solves Eq.(\ref{87a}) relative to
$a_{l}$, the obtained value (the inverse of the r.h.s. Eq.(\ref{47a}))
gives $b_{l}^{-1}$.  In the same manner, one finds the knot value of
$a_{k'}$ on the $k'$-outgoing trajectory as the inverse of the root of
Eq.(\ref{87a}) relative to $b_{k'}$.  The procedure does not pose
calculational problems since the determinant is a linear function of
any of $a$'s or $b$'s.  Eq.(\ref{87a}) represents most concise and
symmetric form of the boundary condition to Eq.(\ref{xra}).

Summarizing, the Green's functions on trajectories linked by a knot is
calculated in the following scheme.  First, one solves Eq.(\ref{xra})
with boundary condition in Eq.(\ref{z7a}) on each of the trajectories
and calculates functions $\phi_{-,m}(x<0)$ and $\phi_{+,k'}(x>0)$; the
parameters $a_{m}$ and $b_{k'}$ in Eq.(\ref{27a}) are then also
known. The next step is to calculate the knot value of $b$'s on the
incoming trajectories and $a$'s on the outgoing ones. This is done by
formulae in Eq.(\ref{47a}) and Eq.(\ref{67a}).  Having obtained the
boundary values, one solves Eq.(\ref{xra}) for $\phi_{+,m}(x<0)$ on
the incoming trajectories and $\phi_{-,k'}(x>0)$ on the outgoing
ones. The 1-point Green's function is then built from $\phi_{\pm}$ by
the recipe in Eq.(\ref{w5a}).

In the Riccati equation technique, one first finds the Andreev
amplitudes $a_{m}(x)$ and $b_{k'}(x)$, $m,k=1,\ldots,N$ from
Eqs.(\ref{hsa},\ref{q5a}). Then, Eqs.(\ref{47a}), and (\ref{67a})
provide the initial value for $b_{m}(x)$ and $a_{k'}(x)$, solutions to
the Riccati equation. The Green's function is then given by
Eq.(\ref{fsa}).

The matching conditions can be also expressed via the transfer matrix
as derived in Section \ref{transs} and in the case of a $N=2$ knot
explained in detail in Section \ref{trans}.

This scheme is also applicable when the trajectories connected by the
knot under consideration may enter other knots.  As a matter of
principle, one assumes that the system under consideration is finite,
and it is surrounded by a ``clean''material where trajectories are
infinite lines without knots.  Then, one solves the problem for the
knots on the boundary and moves inwards towards the knot of interest.
In the one-dimensional topology of the tree with only one path
connecting any two knots, the procedure is unique.

\subsection{$2\times 2$ case}

The most simple case is when the knot mixes two incoming and to two
outgoing trajectories ($N=2$) as e.g. in case of specular reflection
on an interface.  The unitary S-matrix coupling $1$ and $2$ incoming
trajectories to $1'$ and $2'$ outgoing ones (see Fig.\ref{N=2}), may
be taken in the form
\begin{equation}
S= \left(
\begin{array}{lr}
r  &  s\\
- s^{*}  & r^{*}
\end{array}
\right)
\;\; , 
 |r|^{2}=R\; ,\;   |s|^{2} =T \; .
\;\;  R + T =1 \; .
\label{97a}
\end{equation}
Here, $r$ and $s$  are the amplitude of the process $1 \rightarrow 1'$
$2 \rightarrow 1'$, respectively.

Presenting the wave function  on each of the trajectories at the knot as
\[
\psi_{1(2)}= \left(
\begin{array}{c}
   a_{1(2)}\\
   1
\end{array}
\right)
\;\; , \;\;
\psi_{1'(2')}= \left(
\begin{array}{c}
 1  \\
   b_{1'(2')}
\end{array}
\right)
\;\; , \;\;  
\]
the matching condition in Eq.(\ref{87a}) gives the following relation 
between the parameters
\begin{equation}
R (1 - a_{1}b_{1'})(1- a_{2}b_{2'})
+
T(1 - a_{1}b_{2'})(1- a_{2}b_{1'}) =0 \; ,
\label{a8a}
\end{equation}
which serves as the boundary condition for Eq.(\ref{xra}) or Riccati
equation Eq.(\ref{hsa}). 

The usage of it has been explained  
in Section \ref{match}. Reiterating, 
the parameters $a_{1,2}$ ($b_{1',2'}$) in Eq.(\ref{a8a}) are found
from the regular solutions to Eq.(\ref{xra}) or Eq.(\ref{hsa}).
 They are {\it independent} from each other
and the properties of the knot.  The actual meaning of Eq.(\ref{a8a})
is that when it is resolved relative to $a_{1,2}$ ($b_{1',2'}$) the
inverse value gives the initial condition $b_{1,2}(x=0)$
($a_{1',2'}(0)$) {\it i.e.}
\begin{eqnarray}
b_{1}(0)& =  & 
{R  (1- a_{2}b_{2'})b_{1'} + T(1- a_{2}b_{1'})b_{2'}
\over R (1- a_{2}b_{2'}) + T(1- a_{2}b_{1'}) } \; ,   
\rule[-3ex]{0ex}{0ex}
          \label{b8a1}\\   
 a_{1'}(0)& =  & 
{R  (1- a_{2}b_{2'})a_{1} + T(1- a_{1}b_{2'})a_{2}
\over R (1- a_{2}b_{2'}) + T(1- a_{1}b_{2'}) }    \; ,
\label{b8a}
\end{eqnarray}
and the expressions for $b_{2}$ and $a_{2'}$ obtained by the substitution $1
\leftrightarrow 2$.

\subsection{Transfer matrix }\label{transs}

Sometimes it is convenient to consider a pair of trajectories, tag
them to 
$1$ and $1'$, as pieces of a single trajectory (see
Fig.\ref{trans22}).
We assign $x<0$ to the path 1 and $x>0$ to 1'.
Then Eq.(\ref{xra}) is valid for any $x$ excepting the knot point  $x=0$. The knot at
the trajectory $1' \leftarrow 1$ is  included via the $2 \times 2$
transfer matrix
${\cal M}_{1' \leftarrow 1}$:
\begin{equation}
\phi (x= +0)= {\cal M}_{1' \leftarrow 1}\phi (x= -0)
\; , \;
\ofi(x=+0)= \ofi(x=-0)\hat{{\cal M}}_{1'\leftarrow 1}^{-1}
\label{x8a}
\end{equation}
as explained in detail in Appendix Sect. \ref{trans}.  The transfer
matrix is found from the requirements that (i) the matching conditions in
Eq.(\ref{wsa}) are satisfied; (ii) waves on the trajectories other
than 1 and 1' are regular.

Denote  $\phi_{+}(x>0)$  ($\phi_{-}(x<0)$) the solution to Eq.(\ref{xra}) regular at $+
\infty $ ($-\infty $) as in Eq.(\ref{3ra}). 
The transfer matrix allows one to continue the solutions across the knot: 
\begin{equation}
\ofi_{-}(+0)= \ofi_{-}(-0){\cal M}_{1' \leftarrow 1}^{-1}
\;\; , \;\;  
\phi_{+}(-0)= {\cal M}^{-1}_{1' \leftarrow 1}\phi_{+}(+0)
\label{y8a}
\end{equation}
In accordance with Eqs.(\ref{a5a}), and (\ref{s6a}), 
the 1-point Green's function $\gR_{1'}$ on the trajectory 1' at the
knot can be found as 
\[
{1\over 2}\left(1 + \gR_{1'} \right) =
{\phi_{+}(+0) \ofi_{-}(+0)
\over  \ofi_{-}(+0)\phi_{+}(+0)}
\;.
\]
Applying Eq.(\ref{y8a}), one gets from here that
\[
{1\over 2}(1 + \gR_{1'})
=
{\phi_{+}(+0) \ofi_{-}(-0){\cal M}_{1' \leftarrow 1}^{-1}
\over  \ofi_{-}(-0){\cal M}_{1' \leftarrow 1}^{-1}\phi_{+}(+0)
}
\; .
\]
Similarly, for the trajectory 1
\[
{1\over 2}(1 + \gR_{1}) = 
{{\cal M}_{1' \leftarrow 1}^{-1}\phi_{+}(+0)\ofi_{-}(-0)
\over \ofi_{-}(-0){\cal M}_{1' \leftarrow 1}^{-1}\phi_{+}(+0)
}
\]

From here $\gR_{1'}{\cal M}_{1' \leftarrow 1}= {\cal
M}_{1' \leftarrow 1}\gR_{1}$ or
\begin{equation}
\gR_{1'}= 
{\cal M}_{1' \leftarrow 1}
\gR_{1}
{\cal M}_{1' \leftarrow 1}^{-1}
\label{z8a}
\end{equation}

For an arbitrary interface, this relation gives the boundary condition
for the quasiclassical 1-point Green's function.  
With the help of Eq.(\ref{i6a}) or Eq.(\ref{n6a}), the transfer matrix
${\cal M}$ is generally expressed via the Green's function on the
other trajectories coupled by the knot. In the next Section \ref{2x2},
we the explicit expression for the transfer matrix is presented for
the simplest case of 2 in- and 2-out channels.

\subsubsection{ 2x2 case}\label{2x2}
  
In the  most important case of a $ 2\times 2$ knot Fig.\ref{N=2} (e.g. a
specular interface),  the transfer matrix 
can be found usual the general  formula derived in
Sect.\ref{trans}. A more simple way is to make the derivation from the
scratch in a specially selected basis (see Sect.\ref{easy}, for
details).

For the knot with the S-matrix in Eq.(\ref{97a}), the transfer matrix Eq.(\ref{n4a}) and its inverse read 
\begin{equation}
{\cal M}_{1'\leftarrow 1}= {(1+R)\over 2r^{*}} 
\left(1 - {T\over 1+R} \hat{g}_{2' \bullet 2}\right) 
\label{53a} \; ,
\end{equation}
\begin{equation}
{\cal M}_{1'\leftarrow 1}^{-1}= {(1+R)\over 2r} 
\left(1 + {T\over 1+R} \hat{g}_{2' \bullet 2}\right)\; ,
\label{0ab}
\end{equation}
where $\hat{g}_{2'\bullet 2}$ is the normalized ($\hat{g}_{2'\bullet
2}^{2}=1$) ``across-knot'' Green's function. It can  be
presented is different forms.

Its matrix structure is most transparent when $\hat{g}_{2'\bullet 2}$ is written
in a factorized form as
\begin{equation}
{1\over 2}\left(1+ \hat{g}_{2'\bullet 2} \right)
= {1\over N}\phi_{2',+}\, \ofi_{2,-} \;\; , \;\;  
N= \ofi_{2,-}\phi_{2',+}
\label{abb}
\end{equation}
where $\phi_{2',+}$ and $\phi_{2,-}$ are the functions introduced in
Sect.\ref{2point}  taken  at the
point adjacent to the knot on the  trajectory 2' or 2. 
They  do not dependent on   the knot parameters $R$ and $T$.
One may think of $\hat{g}_{2'\bullet 2}$ as a 1-point Green's function
on the virtual trajectory built of the pieces 2 and 2'. 

From Eqs.(\ref{abb}) and (\ref{44a}), one concludes that 
$ (1+\hat{g}_{2'\bullet 2) }
\propto (1+ \gR_{2'})(1+ \gR_{2})$ 

Eq.(\ref{abb}) can written
in terms of the Andreev amplitudes Eq.(\ref{dsa}), as
\[
{1\over 2}\left(1+\hat{g}_{2'\bullet 2} \right)
= {1\over 1- a_{2}b_{2'}}
\left(
\begin{array}{c}
   1    \\
   b_{2'}
\end{array}
\right)
\left(1 , - a_{2}\right)
\] 
or
\begin{equation}
\hat{g}_{2'\bullet 2} = {1\over 1 -a_{2}b_{2'}}
\left(
\begin{array}{lr}
1+ a_{2}b_{2'}& -2 a_{2}\\
 2b_{2'}& - (1 + a_{2}b_{2'})
\end{array}
\right)
\; .
\label{63a}
\end{equation}

The ``across interface'' Green's function can also be written as 
\begin{equation}
\hat{g}_{2'\bullet 2} = {1\over 1 +
{1\over 2}[\hat{g}_{2'},\hat{g}_{2}]_{+}}
\left(\hat{g}_{2'} + \hat{g}_{2} + {1\over
2}[\hat{g}_{2'},\hat{g}_{2}]_{-}
\right)\;\; ,
\label{58a}
\end{equation}
where $\hat{g}_{2,2'}$ are the knot values of the usual 1-point
Green's function on the trajectory 2 and 2'. One should realize that
unlike $\phi_{\pm}$ in Eq.(\ref{abb}) and $a, b$'
 in Eq.(\ref{63a}), both  $ \hat{g}_{2}$ and $ \hat{g}_{2'}$
are modified by the knot scattering, and only their combination 
$\hat{g}_{2'\bullet 2}$ is knot independent.

Using the transfer matrix approach, one can derive the boundary
condition to the Riccati equation on the N=2 knot. Most easily this
can be done using the transfer matrix in Eq.(\ref{cbb}).  Same the
result one can get from Eq.(\ref{a8a}).

We have just presented the boundary condition for the Green's function
on an interface which mixes 2 in-coming and 2 out-going trajectories
(e.g. for a specular interface): the Green's functions on the
interface are linearly related by Eq.(\ref{z8a}) (and the analogous
relation for the channel 2 and 2') where the transfer matrix ${\cal
M}$ and ${\cal M}^{-1}$ can be found from Eqs.(\ref{53a}), (\ref{0ab})
and (\ref{58a}).  Using these relations, one is able to re-derive
Zaitsev's boundary conditions \cite{Zai84} for a specular reflecting
interface.

\section{ Multilayer systems} \label{layer}

The purpose of this section is to show the usage of the general theory
in practical calculations. First we consider simplest geometry that is
a layer deposited on the flat surface of a bulk material with a
partially transparent interface. Together with the totally reflecting
outer surface, there are two coherently reflecting planes.  The other
geometry is a system of two layers of arbitrary thickness in contact,
in which there are three reflecting planes and rather complicated
picture of multiple scattering.

Since out main intention is demonstrate how to use the general
formula, we allow ourselves not to worry about 
the self-consistency of the pair potential. For simplicity, 
we consider the ballistic case $\Sigma_{imp}=0$, and the pair
potentials in the left (l) and right (r) regions  are taken 
constants $\Delta_{l}$ and $\Delta_{r}$.

\subsection{A film}\label{afilm}

The tree-like trajectory near the interface between a layer of
thickness $d_{r}$ and semi-infinite space is shown in
Fig.\ref{film}(a).  To build the tree, one considers a particle coming
along the path (at the angle $\theta $) marked in Fig.\ref{film} by
``1'' which denote both the location and direction.  Due to the
partial reflection, a wave on the trajectory ``4'' is generated. The
waves on the paths ``2'' and ``3'' are generated due to transmission.
The paths ``2'' and ``3'' are the semi-infinite, whereas the
trajectory ``4'' comes to the interface again as ``5''(the total
reflection does not interrupt motion in between ``4'' and
``5''). Again, waves on ``6'' and ``7'' are generated, and the path
continues towards ``9'' {\em etc.}.  The topological structure of the
tree-like trajectory is presented in Fig.\ref{film}(b).

To find two-point Green's function $\gR(x_{1},x_{2})$, one solves
Eq.(\ref{vra}) where the coordinates $x_{1,2}$ correspond now to the
points on the tree Fig. \ref{film}(b) with the understanding that the
tree coordinate $x$ includes information about both the position and
direction of the momentum.  Due to the one-dimensional topology of the
tree, the method described in Sect.\ref{green} is directly applicable.
As before, the 1-point Green's function $\gR(x)$ is given by
Eq.(\ref{04a}).

The matrix $\hat{H}^{R}$ in Eq.(\ref{vra}) is either
\[
\hat{H}^{R}_{l}=
\left(
\begin{array}{lr}
\varepsilon + i \delta  &  \Delta_{l}\\
- \Delta ^{*}_{l}&- \varepsilon   - i \delta 
\end{array}
\right)
\text{ or }
\hat{H}^{R}_{r}=
\left(
\begin{array}{lr}
\varepsilon + i \delta  &  \Delta_{r}\\
- \Delta ^{*}_{r}&- \varepsilon   - i \delta 
\end{array}
\right)
\]
for the tree coordinate $x$ in the left or right regions.
For future references, the free bulk  1-point Green's function in the
left (right) region $\gR_{0,l(r)}$ equals
\[
\gR_{0,l(r)}= {1\over \xi^{R}_{l(r)}}\hat{H}^{R}_{r(l)}
\]
where 
$\xi_{l(r)}^{R} = \sqrt{(\varepsilon + i \delta )^{2} -
|\Delta_{l(r)}|^{2}}$, $\Im\; \xi_{l(r)}^{R} > 0$.

 Considered as a function of $x_{1}$,
$\gR(x_{1},x_{2})$ has a source at $x_{1}= x_{2}$ which
generates waves propagating away from $x_{2}$. The
regularity condition requires that the waves decay when propagating
from the source to  branches of the tree.
The propagation in between the knots is
described by Eqs.(\ref{xra}), or (\ref{z4a}), and  the knots are
incorporated by the matching conditions in Eq.(\ref{wsa}) or their
more advanced version in Eqs.(\ref{87a}),  or  (\ref{x8a}).

Let us first find  1-point Green's function at the tree point $x$  in between
``4'' and ``5''. In accordance with Sect.\ref{green}, one has to find
solutions $\phi_{+}$ which describes the wave spreading from the point
$x$ in the positive direction, and $\phi_{-}$ propagating in the
opposite direction. In the present example, the wave $\phi_{+}$ 
spreads to the paths  ``5'',``6'', ``7'', ``8''\ldots, and
$\phi_{-}$ spreads to ``4'',''3'',''2'',''1''\ldots . 
We chose to think that the particle moves along the ``root''
path ``1''$\rightarrow$``4''$\rightarrow$``5''$\rightarrow $
``8''$\rightarrow$``9''\ldots , and exclude 
 the ``side'' branches ``2'',''3'', ``6'',''7'',
\ldots using the
transfer matrix approach (see Eq.(\ref{y8a})).

Take e.g. the knot where the trajectories ``1-4'' meet (see
Fig.\ref{film}(a)).  The transfer matrix ${\cal M}_{4 \leftarrow 1 }$
can be expressed in accordance with Eq.(\ref{53a}) via the
``across-knot'' Green's function $\hat{g}_{3\bullet 2}$.  In the
present simple case, when ``2'' and ```3'' extend to infinity and
$\hat{H}^{R}$ is same for ``2'' and ``3'', one can conclude from
Eq.(\ref{abb}) or Eq.(\ref{63a}) that $\hat{g}_{3\bullet 2}=
\gR_{0,l}\,$, where $\gR_{0,l}$ is the bulk Green's function in the left
region. 
For any of the identical knots,
the transfer matrix
 reads
\[
{\cal M} = 
{(1+R)\over 2r^{*}}
\left(1 - {T\over 1+R} \gR_{0,l} \right)
\]
where 
$R=|r|^{2}$ and $T=1-R$ 
are the interface  reflection and transmission probabilities.

The functions $\phi_{\pm}(x)$ on the root trajectory,
where $x$ is the coordinated along the root counted from a knot,
 are found with
the help of Eq.(\ref{xra}) supplemented with the boundary condition
connecting the 2-component amplitude leaving the knot $\phi_{out}$
(out = ``4'', ``8'', ``12'', \ldots) via the incoming wave $\phi_{in}$
(in = ``1'', ``5'',''9'', \ldots)
\[
\phi_{out} = {\cal M} \phi_{in} .
\]
In the present case, when the free motion on the root trajectory is
perturbed by the equidistant knots, one can use the method developed
in Sect.\ref{append.period} for periodic potentials.  The period of
the structure is $2D_{\theta }$, $D_{\theta }= d_{r}/\cos\theta $
where $\theta $ is the angle between the direction of the momentum and
the perpendicular to the interface.

The functions $\phi_{\pm}(x)$ are eigenfunctions of the evolution
operator $\hat{U}_{2D_{\theta}}(x)$ generating the translation by the
period $x \rightarrow x + 2D_{\theta }$ (see
Sect.\ref{append.period}).  The free evolution operator
$\hat{U}^{(r)}(x+ x_{0}, x_{0})$ in the right region is
\[
\hat{U}^{(r)}(x+ x_{0}, x_{0}) = e^{{i\over v} \hat{H}_{r}^{R}x}
= \cos( {\xi^{R}x\over v}) + i \gR_{0,r} \sin( {\xi^{R}x\over v}),
\]
$\hat{g}_{0,r}^{R}$ being the bulk Green's function.

The full evolution operator $\hat{U}_{2D_{\theta}}(x)$ reads 
\[
\hat{U}_{0}(x)= 
A \exp\left({i \xi_{r}^{R}\over v} \hat{g}_{0,r}^{R}x\right)
\left(1 - {T\over 1+R} \; \gR_{0,l}\right)
\exp\left({i \xi_{r}^{R}\over v} \hat{g}_{0,r}^{R}(2D_{\theta } -x)\right)
\, ,
\]
where $A= {(1+R)\over 2r^{*}}$.
Finding the two eigenfunctions of this matrix, one knows 
$\phi_{\pm}(x)$ and, therefore, the full 2-point Green's function from
Eq.(\ref{7ra}). 

As explained in Sect.\ref{append.period}, the 1-point Green's function
can be extracted from $\hat{U}_{2D_{\theta}}(x)$  by purely algebraic
transformations.
The Green's function for the direction of the momentum $(\bbox{p})_{z}=
p_{F} \cos \theta $ 
 at the distance from the interface $z$
($z>0$ in the right region)
 reads from Eq.(\ref{y7a}) 
\begin{equation}
\gR(z, \theta ) = 
\left.
\Format{2}{
\exp\left({i \xi_{r}^{R}\over v} \hat{g}_{0,r}^{R}x\right)
\left(1 - {T\over 1+R} \; \gR_{0,l}\right)
\exp\left({i \xi_{r}^{R}\over v} \hat{g}_{0,r}^{R}(2D_{\theta } -x)\right)
}
\right|_{x = {z\over \cos\theta }}
\label{ibb}
\end{equation}
where the ``formatting'' operation $\Format{0.7}{\ldots}$ is defined in
Eq.(\ref{y7a}). The ``formatting'' can be performed analytically but
the result looks rather awkward and hardly any information can be
extracted from it without a computer.  On the other hand, the
``formatting'' operation is easily implemented numerically, and for
this reason we leave as final the expression for Green's function
  in Eq.(\ref{ibb}).

Consider now the left region and the knot ``1''- ``4'' in
Fig.\ref{film}(a).  The left region Green's functions are those on
trajectories ``2'' and ``3''. To apply formula in Sect. \ref{trans}
and one should substitute 1 for ``2'' and 1' for ``3''.  Since
trajectories ``2'' and ``3'' are semi-infinite, the combination
$\phi_{+}(+0) \ofi_{-}(-0)$ is proportional to the bulk value $(1 +
\gR_{0,l})$. The transfer matrix ${\cal M}_{4 \leftarrow 1}$ contains
the across-knot Green's function $\hat{g}_{4\bullet 1}$ analogously to
Eq.(\ref{58a}). It is easy to see that $\hat{g}_{4\bullet 1}$ equals
to just found $\gR(z=+0, \theta )$.  Therefore, the Green's function
on the left side of the interface is 
\[
\gR(z=-0, \theta )=
\Format{2}{
(1 + \gR_{0,l}) 
\left(1 + {T\over 1+R} \;\gR(z=+0, \theta )\right)
}
\]
At other points  in the left region ($z<0$),
 the Green's function is found
with the help of the free evolution operator,
\begin{equation}
\gR(z, \theta )= 
\left.
\exp\left({i \xi_{l}^{R}\over v} \gR_{0,l}x\right)
\gR(z=-0, \theta )
\exp\left(-{i \xi_{l}^{R}\over v} \gR_{0,l}x\right)
\right|_{x = |z/\cos\theta |}
\label{mbb}
\end{equation}

In Fig.\ref{layer.dos}, we show the density of states on the film side
of the interface, {\it i.e.}  $\Im \; \gR(z=+0, \theta )$
Eq.(\ref{ibb}), for $D_{\theta }= v/ |\Delta_{l}|$ and the pair
potential in the left and right parts of different signs, $\Delta_{l}=
-\Delta_{r}$; the curves parameters differ in the reflectivity $R$
increasing from zero in Fig.\ref{layer.dos}(a) to R=0.9 in
Fig.\ref{layer.dos}(d).

When $R=0$, one sees in
Fig.\ref{layer.dos}(a)   two (zero width) peaks in the  gap
region $|\varepsilon |< |\Delta |$.  The peaks are due to
the bound   states well-known
known in the theory of anisotropic superconductors 
\cite{FogRaiSau97} 
(see also Sect.\ref{rough}). 
The $\varepsilon =0$  bound states exist near the trajectory point where the
phase of $\Delta$ changes abruptly by $\pi $. When the thickness
$d_{r}$ is finite, the levels are
at a finite energy \cite{FauBelBla99}
 due to the overlap of the wave functions 
(e.g. of the states on the  
``2''-``4'' and  ``5''-``7'' paths in Fig. \ref{film}(a)) and the level
repulsion. 
The overlap of the separated in space levels and,
therefore, the level splitting are exponentially small when $D_{\theta
}$ is large.

When $R$ is finite, the splitting increases. 
First, the reflection gives rise to the on-knot overlap of the levels belonging to
the the same knot, e.g. the  ``2''-``4'' and ``1''- ``3''
levels". By this mechanism, the level is split to  $\pm \sqrt{R}|\Delta |$
(cf. Eq.(\ref{x7a})). Second, the on-knot overlap in combination
with the next neighbour overlap discussed earlier, mixes
together all the bound states and
transforms the discreet levels into bands. This behaviour is clearly seen in
Fig.\ref{layer.dos}(b)-(d).

\subsection{Sandwich }\label{sand}

In this Section we consider a more general case when the 
left region is a finite layer of thickness $d_{l}$. As  previously,
the order parameter is assumed to be constant in the layers.

The typical tree-like trajectory formed by multiple reflections on the
outer surfaces and the interface, is shown in
Fig.\ref{sand+tree}(a). As in Fig.\ref{film}, the numbers tag the
coordinate on the trajectory. Topological structure of (a fragment of)
the tree is shown in Fig.\ref{sand+tree}(b); the tagging in same is in
Fig.\ref{sand+tree}(a).  The centre of the tree is (arbitrarily)
chosen at the ``5''-``8'' knot; the tree structure looks same if
viewed from different knots.  The pieces of the tree with the arrows
in the horizontal direction correspond to the the left layer, and
points on the vertical lines belong to the right layer. Generally, the
tree-like trajectory covers (almost) all space but remains
nevertheless topologically one-dimensional: The features discussed
before are clearly seen here  that is (i) if a line of
the tree is cut, two disconnected pieces are produced or,
equivalently, (ii) there is no closed loops on the tree.

First we calculate the knot values of the Green's functions, for the
central knot ``5''-``8''. Other knots are equivalent to the central
knot. On both horizontal and vertical branches in
Fig. \ref{sand+tree}, the arrays of knots are periodical, separated by
$2D_{l,\theta }$, $D_{l,\theta }= d_{l}/\cos\theta $ for the
horizontal branches (the left layer) and $2D_{r,\theta }$,
$D_{r,\theta }= d_{r}/\cos\theta $, on the vertical branches (the
right layer).

As in the previous section (see Sect. \ref{append.period} for prove),
the 1-point Green's function at ``5'', $\gR_{5}$, is simply
related to the evolution operator $\hat{U_{9 \leftarrow 5}}$ advancing
the wave function at ``5'' to the periodically equivalent point ``9''
(see Fig. \ref{sand+tree})(b). Crossing the knot from ``5'' to ``8''
with the help of the transfer  matrix, ${\cal M}_{\downarrow }$,
 build analogously to  Eq.(\ref{53a}),
\[
{\cal M}_{\downarrow }= 
{(1+R)\over 2r^{*}}
\left(1 - {T\over 1+R} \hat{g}_{7 \bullet 6}\right)
 \; ,
\]
and moving 
from ``8'' to ``9'' by
 $\exp(2i D_{\theta ,r}\gR_{0,r})$, one get  $\hat{U_{9
\leftarrow 5}}$ as the ordered product of the two matrices. 
The same matrices but multiplied in the different order, give the
evolution operator $\hat{U_{8 \leftarrow 3}}$ and, therefore 
$\gR_{8}$.

Changing notation in Eq.(\ref{58a}) and collecting formulae together, 
one gets
\begin{mathletters}\label{vert}
\begin{eqnarray}
\hat{g}_{7 \bullet 6}     & =  &
{1\over 1 +
{1\over 2}[\hat{g}_{7},\hat{g}_{6}]_{+}}
\left(\hat{g}_{7} + \hat{g}_{6} + {1\over
2}[\hat{g}_{7},\hat{g}_{6}]_{-}
\right)
\rule[-3ex]{0ex}{0ex}
\label{qbb1}\\   
\gR_{8}     & =  & 
\Format{2}{
\left(1 - {T\over 1+R} \hat{g}_{7 \bullet 6}\right)
\exp(2i D_{\theta ,r}\gR_{0,r}/v)
}
\rule[-3ex]{0ex}{0ex}
\label{qbb2}\\   
\gR_{5}     & =  & 
\Format{2}{
\exp(2i D_{\theta ,r}\gR_{0,r}/v)
\left(1 - {T\over 1+R} \hat{g}_{7 \bullet 6}\right)
}
\; .
\label{qbb}
\end{eqnarray}
 \end{mathletters}
These equations allow one to find the knot values of the Green's
function in the right region via the left region counterparts.

In the same way one can 
derive expressions where $\gR_{6,7}$ are related
to $\gR_{5,8}$
\begin{mathletters}\label{horiz} 
\begin{eqnarray}
\hat{g}_{8 \bullet 5}     & =  &
{1\over 1 +
{1\over 2}[\hat{g}_{8},\hat{g}_{5}]_{+}}
\left(\hat{g}_{8} + \hat{g}_{5} + {1\over
2}[\hat{g}_{8},\hat{g}_{5}]_{-}
\right)
\rule[-3ex]{0ex}{0ex}
\label{rbb1}\\   
\gR_{7}     & =  & 
\Format{2}{
\left(1 - {T\over 1+R} \hat{g}_{8 \bullet 6}\right)
\exp(2i D_{\theta ,l}\gR_{0,l}/v)
}
\rule[-3ex]{0ex}{0ex}
\label{rbb2}\\   
\gR_{6}     & =  & 
\Format{2}{
\exp(2i D_{\theta ,l}\gR_{0,l}/v)
\left(1 - {T\over 1+R} \hat{g}_{8 \bullet 5}\right)
}
\; .
\label{rbb}
\end{eqnarray}
\end{mathletters}

Eqs.(\ref{vert}) and (\ref{horiz}) allow one to find iteratively the
knot values of the Green's function. Unless the reflection $R$ is too
small, the iterations converge rather fast. For almost transparent
interfaces, $R\ll 1$, a slightly different procedure is more
efficient: as the periods, one chooses the paths like ``4''
$\rightarrow $``5'' $\rightarrow $ ``7'' $\rightarrow $ ``14''
$\rightarrow $ ``16''.

Given the knot values, the Green's function at other points
can be calculated by formulae analogous to Eq.(\ref{mbb}).

Fig.\ref{sandwich} shows the trajectory resolved density of states at
the interface, $\Im \; \gR_{z=0, \theta }$ for the case when
the $\Delta_{l}= - \Delta_{r}$ and the layers of equal thickness
$D_{l,\theta }=D_{r,\theta }= v/|\Delta_{l}|$.

As expected, the sandwich with a transparent interface, $R=0$, has a
considerable spectral weight at low energies which is represented by
the band centred at $\varepsilon =0$ (see Fig.\ref{sandwich}(a)). The
overall picture is very different from the BCS density of states: the
spectrum is given by well-defined bands with strong edge
singularities.  As in case of a film, the reflection splits the
$\varepsilon =0$ bound states, and the bands move towards higher
energies. When the reflectivity is as low as 0.1 (see
Fig.\ref{sandwich}(b)), there is no states at, and in the vicinity of
$\varepsilon =0$.  The forbidden bands become more narrow, and the
edge singularities become smoother.  From Fig.\ref{sandwich}(c) and
(d), one sees that for $R\gtrsim 0.5$ the states are pushed to the
energies $\gtrsim \Delta $.

In the next section, we use these results to evaluate the ``superfluid
density'', an observable sensitive to the shape of the density of
states.

\subsubsection{Superfluid density}\label{ros}

In this section we calculate  $\rho_{s}$, a
parameter which controls the current density $\bbox{j}$ induced by a
weak spatially homogeneous static vector potential, $\bbox{A}$,
\[
\bbox{j} = - \rho_{s} {c\over 4\pi } {1\over \lambda_{L}^{2}}
\bbox{A}
\; ,
\]
$ \lambda_{L}$ being the (bulk)  London penetration depth at zero
temperature. In the two-fluid lexicon, $\rho_{s}$ is 
the ``superfluid density'' or
 the ``fraction of superconducting electrons''.

In the present case of a two-layer system, the local current induced
by in-plane homogeneous vector potential is $z$-dependent, being
proportional to the local density of states. The total current through
the layers is proportional to the average,
\begin{equation}
\rho_{s}= {1\over d_{l}+d_{r}} 
\int\limits_{-d_{l}}^{d_{r}}dz\; \rho_{s}(z)
\;\; , \;\;
\rho_{s}(z)= 1 -\int\limits_{-\infty}^{\infty} d \varepsilon \;
 \left(- {\partial
f_{0}\over{\partial\varepsilon}}\right) \nu(\varepsilon ,z)
\label{vbb}
\end{equation}
where  $f_{0}$ is the Fermi function, 
and $\nu(\varepsilon ,z)$ is the local density of states,
\[
\nu(\varepsilon ,z) = Re\; \int {d \Omega_{\bbox{n}}\over 4
\pi } 
\left(\gR(\varepsilon, \bbox{n}, z) \right)_{11}
\; .
\]

The  averaged superfluid density  $\rho_{s}$ in Eq.(\ref{vbb}) is
conveniently written as
\begin{equation}
\rho_{s}=
 1 -\int\limits_{-\infty}^{\infty} d \varepsilon \;
 \left(- {\partial
f_{0}\over{\partial\varepsilon}}\right) \bar{\nu}(\varepsilon) 
\;\; , \;\;  
 \bar{\nu}(\varepsilon) \equiv {d_{l} \bar{\nu}_{l}(\varepsilon) + d_{r}
 \bar{\nu}_{r}(\varepsilon)\over  d_{l} + d_{r}}
\label{wbb}
\end{equation}
where $ \bar{\nu}_{l,r}(\varepsilon)$ is the averaged density of
states in the left ($l$) and right ($r$) layers.

To calculate $ \bar{\nu}_{l,r}(\varepsilon)$, one finds the Green's
function, as explained in the previous section, and perform
integrations with respect to the coordinate $z$ and the direction
$\bbox{n}$.  The spatial dependence, found from the knot values by
formulae analogous to Eq.(\ref{mbb}), is simple and the
$z-$integration can be done analytically.  The averaged density of
states in the right region reads
\begin{eqnarray}
 \bar{\nu}_{r}(\varepsilon)    & =  &
{1\over 2}{\rm Sp}\,\tau_{z}\;
 \Re\; \int\limits_{0}^{1} d \mu
\; \,\Big( 
\gR_{0,r}[\gR_{0,r},\gR_{5}]_{+}
\nonumber\\        
&   &+ 
{\sin 2\gamma \over 2 \gamma }
\gR_{0,r}[\gR_{0,r},\gR_{5}]_{-}
+i
\left({1 - \cos 2 \gamma \over 2 \gamma }  \right)
 [\gR_{0,r},\gR_{5}]_{-}
\Big)_{\varepsilon ,\mu }
\label{ybb}
\end{eqnarray}
where, $\mu = \cos \theta $, $\gamma_{\varepsilon , \mu }\equiv 2d_{r}
\xi^{R}_{r}/ \mu v$, $\gR_{0,r}$ is the bulk Green's function,
and $\gR_{5}$ is the knot value of the Green's function on the
tree corresponding to the angle $\theta $ (see
previous Section).
After the substitution, $r \rightarrow l$ and $\gR_{5}
\rightarrow \gR_{6}$, Eq.(\ref{ybb}) gives $\bar{\nu }_{l}$.

The integration with respect to $\mu $ in Eq.(\ref{ybb}) and
$\varepsilon $ in Eq.(\ref{wbb}) can be performed only numerically.
The integration in Eq.(\ref{wbb}) along the real $\varepsilon $ axes
may be slowly converging due to the band edge singularities; for
better convergence, one may integrate along line $\Im \; \varepsilon =
i {T\pi\over 2}$ or transform the integral to the Matsubara sum.

We evaluated numerically  the superfluid density
for a sandwich with equal thickness of the layers $d_{l,r}=
v/|\Delta_{l,r}|$ and the $\pi $ differences in the order parameter
phase $\Delta_{l}= - \Delta_{r}$.  In Fig.\ref{rhoss}, the superfluid
density as a function of temperature is shown for different
reflectivity $R$.

The curve for $R=0$ shows large negative $\rho_{s}$ at low
temperatures which would lead to amplification of the applied magnetic
field rather than the Meissner screening.  This feature is due to the
large low energy spectral weight seen in Fig.\ref{sandwich}(a).
Therefore, our data support the recent idea put forward by Fauchere,
Belzig, and Blatter \cite{FauBelBla99} about the paramagnetic
instability near the surface where the order parameter changes its
sign.  However, one sees in Fig.\ref{sandwich} that the effect is very
sensitive to the presence of the partially reflective interface:
reflection with the probability as low as 4 percent makes $\rho_{s}$
positive at any temperature.

\section{Conclusions}\label{concl}

In this paper we have reconsidered the part of the quasiclassical
theory of superconductivity which concerns interfaces between
superconductors (SIS) or a normal metal and a superconductors (NIS).
Since the interface violates the condition of applicability of the
quasiclassical approximation, the reflection and transmission
processes must be included via a boundary condition. In the approach
taken in the paper, the master boundary condition in Eq.(\ref{wsa}) is
formulated for the effective wave functions factorizing the 2-point
Green's function. In the boundary condition, the two-component
amplitudes in N in-coming and N out-going channels are related to each
other via the S-matrix. The latter is sensitive to microscopic details
of the interface and is considered as an input in the quasiclassical
theory.  The theory is equally applicable to specular interfaces
($N=2$), as well as to the many channel case which models a rough
surface or interface.  In Sections \ref{andreev}, \ref{match}, and
\ref{transs}, the
master boundary condition is reformulated in various forms, suitable
for the one or the other application.

In Sect.\ref{andreev}, we have presented a general solution to the
ballistic problem of the scattering of electron-and hole-like {\em
excitation}.  This result extends the theory of the NIS interface
\cite{She80b} to the many channel situation; SIS case is also
included.  As in Ref.\cite{She80b}, the solution is general in the
sense that it expresses the full amplitude of the multiple processes
of the Andreev electron $\leftrightarrow$ hole conversion and ordinary
scattering via the amplitudes of the elementary processes. By this,
the problem is split into independent and more simple problems. The
theory of multi-channel bound states is also considered.  The
formulation which operates with excitations rather than bare
particles, is especially convenient for the kinetic theory in the
framework of the Boltzmann-type equations, for which it provides the
boundary condition for the distribution function of the excitations
\cite{She80c,ShelFtt}.

For a general case {\it i.e.}  when the disorder and inelastic
collisions are allowed, the boundary value of the 2-component wave
functions $\phi = {u\choose v}$ factorizing the trajectory Green's
function are found in Sect.\ref{match}.  Since the mean field
equations are linear, this result can be recast as the boundary
condition for the Andreev amplitudes $u/v$ of the Riccati equation
approach.  In a most compact and symmetric form, the boundary
condition is given by Eq.(\ref{87a}). For the specular interface, the
boundary condition for the Riccati equation is given by
Eq.(\ref{a8a}), or explicitly by Eqs.(\ref{b8a1}), and (\ref{b8a}).

One more form of the boundary condition is presented in
Sect.\ref{transs}, where the expression for the transfer matrix is
derived. The transfer matrix, which couples the wave functions or the
1-point Green's functions on the chosen pair of in- and out-channels,
absorbs information about all other $2(N-1)$ channels.  This
modification of the boundary condition is convenient when one solves
the Eilenberger equations for the 1-point Green's function. In the
simplest 2-channel case (specular reflection), this boundary condition
reproduces Zaitsev's results \cite{Zai84}.  The new form seems to be
more flexible and convenient.

For the derivation, we use the technique of the 2-point Green's
function.  In our opinion, the technique provides an adequate language
to discuss the semiclassical physics in superconductors which we
qualitatively considered in Section \ref{intro}. The 2-point Green's
function gives a full description of the coherent propagation of
electron and hole along a common classical path.  In spite of the fact
that observables can be expressed via the 1-point Green's function
only, the language of the quasiclassical 2-point Green's function on
is not redundant: Offering a physically transparent formalism, it is
free from some uniqueness problems which plague the standard
``$\xi$-integrated'' formulation.  Note also that with all possible
simplifications already done, the quasiclassical 2-point Green's
function obeys Eqs.(\ref{vra1}), and (\ref{vra}) which, unlike the
Eilenberger equations, have a familiar form of an equation for a
propagator.  Therefore, one may directly apply the intuition and
experience gained in other fields of the quantum theory.

Another attractive feature of the 2-point Green's function technique
is that it allows one to define effective wave functions. The latter
factorize the Green's function averaged with respect to disorder or
phonons.  Although these ``wave functions'' have usual quantum
mechanical meaning only in ballistic case, it seems to be advantageous
that one may use the unified language of trajectories and wave
functions discussing both the ballistic motion and the propagation in
the presence of disorder or inelastic collisions.

The effective wave function, $ \phi = {u \choose v}$ obeys the linear
Andreev-type equation Eq.(\ref{xra}).  There is a variety of methods
one can chose to solve the system of two linear differential equations
for $u$ and $v$. One of them is to derive the equation for the ratio
$u/v$ which turns out to be the Riccati equation suggested in
\cite{SchMak95,Sch98}.  As the logarithmic derivative $\psi '/ \psi $
in the usual Schr\"odinger equation does, the choice of the ratio
$u/v$ has the indisputable practical advantage which is due to
insensitivity of the ratio to the normalisation of $\phi $.  The
Riccati equation approach which has proven to be very convenient and
efficient for numerics, finds rather natural physical interpretation
in the 2-point Green's function technique of the present paper.  (For
the latest development of the Riccati equation approach including the
interface boundary condition see e-preprint of M. Eschrig
\cite{Esc99}.)

An important part of this paper is the understanding that the
classical trajectory transforms to a topologically 1-dimensional
simply connected tree in the case of many interfaces and/or
boundaries. The extended arguments in favour of this point of view
have been presented in Section \ref{intro}.  Although, this assertion
may look wrong in simple idealized geometries, like e.g., a sandwich
with strictly parallel outer and the interface planes Fig.\ref{loop1},
we argue that small deviations from the perfection eliminate
accidental crossings of trajectories (as in non-integrable
billiards).  In our opinion, the difficulties with the quasiclassical
theory encountered in \cite{AshAoyHar89,Nag98} are due to the fact
that some interference contributions survive the procedure of the
integration with respect to the layer thickness: Indeed, rigid
variations of the layer thickness do not eliminate all the loops.  We
believe that some roughness, larger then the Fermi wave length but
small and invisible on the quasiclassical scale, will restore the
quasiclassical results.

To show the new theory in action, we solve in Sect.\ref{layer} two
simple problems: (i) a film separated from a bulk material by a
partially transparent interface; (ii) a two layer system with
arbitrary transparent interface. (The latter was classified in
\cite{Nag98} as quasiclassically unsolvable.)  Motivated by recent
ideas about the origin of the paramagnetic effect \cite{Mota}, we
evaluate the density of states and the superfluid density when the
phase of the order parameters in the layers differs in $\pi $, a
scenario of paramagnetic instability suggested in
\cite{FauBelBla99}. Our results confirm the very possibility that the
superfluid density $\rho_{s}$ may be negative (Meissner {\em
``anti-screening''}) but we observe also that $\rho_{s}$ is strongly
affected by reflection on the interface: when the probability of the
reflection $R> 0.04$, the Meissner {\em screening} is restored.  The
implications of these results for a realistic theory of the
paramagnetic instability requires further studies.

\acknowledgments

We are thankful to W. Belzig and C. Bruder for discussions, and to
D. Rainer for very useful comments.  This study begun during the stay
of one of us (A.S) at the Institut f\"ur Theorie der Kondensierten
Materie, Universit\"at Karlsruhe, and A.S.  would like to thank all
the stuff for hospitality and der Deutschen Forschungsgemeinschaft
(SFB 195) for support.  In part this work was supported by the Swedish
Natural Science Research Council.

\appendix

\section{Advanced Green's functions}\label{ga}

The advanced Green's function $g^{A}(x_{1},x_{2})$ is constructed in 
the same manner as the retarded one: One finds $\phi_{1,2}$ from
Eq.(\ref{xra}) with $\hat{H}^{R}$ substitutes for $\hat{H}^{A}$ and
builds the Green's function  as in Eqs.(\ref{7ra}), (\ref{44a}) and
(\ref{fsa}). 

Combining Eqs.(\ref{i5a}) and (\ref{zra}), one can see that the
$\hat{\tau}_{x}\left(\phi^{R} \right)^{*}$ with $\phi^{R}$ from
Eq.(\ref{xra}) satisfies the corresponding equation in the A-case.
Then, the normalized Eq.(\ref{8ra}) solutions are related to each
other as
\begin{equation}
\phi_{+}^{A}=
i\hat{\tau}_{x}\left(\phi_{+}^{R} \right)^{*}
\;\; , \;\;
\phi_{-}^{A}=
i\hat{\tau}_{x}\left(\phi_{-}^{R} \right)^{*} \;\; .
\label{j5a}
\end{equation}

The Andreev amplitudes $a$ and $b$ Eq.(\ref{dsa})
are related now as
\[
a^{A} = 1/\left( a^{R} \right)^{*}
\;\; , \;\;  
b^{A} = 1/\left( b^{R} \right)^{*} \;,
\]
and Green's functions as 
\[
\gR(x_{1},x_{2})=
\hat{\tau}_{x}
\hat{g}^{A*}(x_{1},x_{2})
\hat{\tau}_{x}
\]

For future references,
the symmetry in the 1-point Green's functions is given by
the following well-known relations 
(
$
\bbox{\varepsilon }= (\varepsilon, \bbox{n}) ,
\bbox{\varepsilon }^{*}= (\varepsilon^{*}, \bbox{n})
$) 
\[
\gR_{\bbox{\varepsilon }}(\bbox{r}) = - \hat{\tau}_{z}
\left(\hat{g}^{A}_{\bbox{\varepsilon }^{*}}(\bbox{r}) \right)^{\dagger}
\hat{\tau}_{z}
\;\; , \;\;  
\gR_{\bbox{\varepsilon }}(\bbox{r}) = 
\left(\gR_{-\bbox{\varepsilon }^{*}}(\bbox{r}) \right)^{\dagger}
\]
The first of them follows from Eq.(\ref{j5a}), and the second one
reflects the symmetry 
$\hat{H}^{R}_{\bbox{\varepsilon }}= - 
\hat{\tau}_{z}\hat{H}^{A}_{-\bbox{\varepsilon }}\hat{\tau}_{z}$.
  
\section{Evolution in periodic potential}
\label{append.period}

To prove validity of Eq.(\ref{u5a}),
one first solves the $2\times 2$ eigenvalue problem 
\[
\hat{U}_{L}(x)\psi(x)  = \gamma \psi(x) \; .
\]
and finds the eigenfunctions $\psi_{1,2}$
(with $x$ as a parameter)
 and  the eigenvalues
$\gamma_{1,2}$.  
It follows from the conservation of normalization in Eq.(\ref{1ra}) that
${\rm Det} \; U =1$, and, therefore,
\[
\gamma_{1}\gamma_{2}=1 \; .
\]
Denote  $\gamma_{1}$ 
the eigenvalue for which  $|\gamma_{1}|<1$,
\footnote{
When considering $\gR$,
the variable $\varepsilon $
 has a finite imaginary part and the matrix $\hat{H}^{R}$
is not Hermitian.  Then, the evolution matrix is   not unitary and 
$|\gamma_{1,2}| \neq 1$.
}  
and normalize the eigenfunctions to satisfy
 $\ops_{2}\psi_{1}=1$.  It is clear now, that 
$\psi_{1}(x)$  continued along the trajectory with
the help of the evolution matrix $\hat{U}_{L}(x)$ gives the solution
denoted in Eq.(\ref{3ra}) as $\phi_{+}(x)$: 
Indeed, it satisfies Eq.(\ref{xra}) and decays
at $x \rightarrow \infty $ as 
$ \gamma_{1}^{x/L}$.  By the
same argument, $\phi_{-}=\psi_{2}$. From Eq.(\ref{w5a}), the Green
function now reads  
\[ 
\gR = \psi_{1} \ops_{2} + \psi_{2}
\ops_{1} \; .  
\]

Seeing that the evolution $2 \times 2$ matrix can be expanded in
its normalized eigenfunctions as
\[
\hat{U}_{L}(x) = 
{1\over 2}(\gamma_{1}+ \gamma_{2}) \hat{1} + 
{1\over 2}(\gamma_{1}- \gamma_{2})\left(\psi_{1} \ops_{2} 
+ \psi_{2}\ops_{1} \right)
\; ,
\]
the traceless part of $\hat{U}_{L}(x)$ is proportional to $\gR$.
The normalization condition fixes the proportionality coefficient, and
one comes to Eq.(\ref{u5a}).

To build the evolution matrix, one may use the following procedure. 
First consider two fundamental solutions to Eq.(\ref{xra}), 
$\psi_{I}$ and $\psi_{II}$, 
which satisfy the following  boundary conditions
\[
\psi_{I}(x)= \Hcolumn{1}{0}
\;\; , \;\;  
\psi_{II}(x+L)= \Hcolumn{0}{1} \; .
\]
and find  $\psi_{I}(x+L)$ and
$\psi_{II}(x)$,
\[
\psi_{I}(x+L) = e^{i \Phi_{L}(x)} \Hcolumn{1}{\alpha_{L}(x)}
\;\; , \;\;  
\psi_{II}(x) = e^{i \Phi_{L}(x)} \Hcolumn{\beta_{L}(x)}{1}\; .
\]
The exponential factor is same for $\psi_{I}$ and $\psi_{II}$ as
required by the conservation of the normalization in Eq.(\ref{1ra}). 
The parameters $\alpha_{L}(x)$, $\beta_{L}(x)$, and $\Phi_{L}(x)$ can
be  calculated conveniently in the Riccati equation  technique.

Denote $\alpha_{0}(x;x_{0})$ solution to 
Eq.(\ref{hsa}) with the boundary condition
$\alpha_{0}(x=x_{0};x_{0})=0$; then (see Eq.(\ref{jsa}))
\[
\alpha_{L}(x) = \alpha_{0}(x+L;x)\;\; , \;\;  
\Phi_{L}(x) = \int\limits_{x}^{x+L} d x' \; \left(
\varepsilon^{R}(x') + \Delta^{R}(x')\, \alpha(x'; x)
 \right) 
\]

Similarly, 
\[
\beta_{L}(x) = \beta_{0}(x, x+L)
\]
where $\left(\beta_{0}(x, x_{0} \right)^{-1}$ is the solution to 
Eq.(\ref{hsa}) with the boundary condition 
$\beta_{0}(x=x_{0}, x_{0})=0$.

Building the evolution matrix from the fundamental solutions, one gets
\[
\hat{U}_{L}(x)= e^{i \Phi_{L}(x)}
\left(
\begin{array}{lr}
1   &  -\beta_{L}(x)  \\
 \alpha _{L}(x)\;\;  & e^{-2 i \Phi_{L}(x)} - \alpha _{L}(x)\beta_{L}(x)
\end{array}
\right)
\]

The traceless part of it, $\hat{U}^{\prime}_{L}(x)$, 
reads
\[
\hat{U}^{\prime}_{L}(x)= 
{1\over 2} e^{i \Phi_{L}(x)}
\left(
\begin{array}{lr}
1 -e^{-2 i \Phi_{L}(x)} + \alpha _{L}(x)\beta_{L}(x)  &  -2\beta_{L}(x) \\
 2\alpha _{L}(x)\;\;  & -1 + e^{-2 i \Phi_{L}(x)} 
- \alpha _{L}(x)\beta_{L}(x)
\end{array}
\right) \; .
\]
Up to the normalization factor, this matrix is equal to $\gR(x)$.

\section{Formal solution: I}\label{formal}

Here, we analyze some formal linear algebra aspects of the matching
conditions in Eq.(\ref{wsa}).

Generally, the wave functions may be presented in the following form:
\begin{equation}
\psi_{k'} = A_{k'} \left(
\begin{array}{l}
1    \\
 \mu_{k'}
\end{array}
\right)
\;\; , \;\;  
\psi_{i} = B_{i} \left(
\begin{array}{l}
\nu_{i}\\
 1  
\end{array}
\right)
\label{85a}
\end{equation}

Denote $|X \rangle $ the column with elements $X_{1}, \ldots , X_{N}$
or $X_{1'},\ldots,X_{N'}$.  One obtains from Eqs.(\ref{wsa}), and
(\ref{zsa})
\begin{equation}
|A \rangle = \hat{S}\hat{\nu}|B \rangle
\;\; , \;\;    
|B \rangle = \hat{S}^{\dagger}\hat{\mu}|A \rangle  \; .
\label{95a}
\end{equation}
where $\hat{\mu }$ and $\hat{\nu }$ are diagonal matrices
$N\times N$ with 
$(\hat{\mu })_{k'k'}= \mu_{k'}$ and 
$(\hat{\nu  })_{kk}= \nu_{k}$.

The two equalities in Eq.(\ref{95a}) are compatible only if 
\begin{equation}
{\cal D}(\{\nu\},\{\mu\})\equiv 
\det
\left|
\left|
1 - 
\hat{S}\hat{\nu} \hat{S}^{\dagger}\hat{\mu}
\right|
\right|
 = 0 \; ,
\label{05a}
\end{equation}
As expected, the parameters $\mu $'s and $\nu$'s are not independent:
Eq.(\ref{05a}) gives a relation among them, which is {\it linear} in
each of the parameters (see Section \ref{det}), making it possible to
express one of the $\mu $'s or $\nu$'s through all others.

For instance, one may give any values to $\mu $'s and $\nu$'s in all
channels excepting the $l-$th incoming one. Then $\nu_{l}$ is fixed by
Eq.(\ref{05a}),
\begin{equation}
\nu_{l}^{-1} = \langle l| 
\hat{S}^{\dagger}\hat{\mu} \hat{\bbox{S}_{l}}
|l \rangle 
\; .
\label{b6a}
\end{equation}
where 
\begin{equation}
\hat{\bbox{S}}_{l} = 
\left(\hat{S}^{\dagger} - \hat{\nu}^{(l)}
\hat{S}^{\dagger}\hat{\mu}  \right)^{-1}\;\; ,
\label{c6a}
\end{equation}
and $\hat{\nu}^{(l)}$ denotes the matrix which differs from
$\hat{\nu}$ only in that the element $(\hat{\nu}^{(l)})_{ll}=0$; here
and below $\langle i | \hat{Q}| j\rangle \equiv Q_{ij}$ .

From Eq.(\ref{95a}) one finds now the coefficients $A$'s and $B$'s,
they are proportional to one of them, say $B_{l}$.
It is convenient to put $B_{l}= C \nu_{l}^{-1}$.
In this way, one gets the solution to the matching conditions in
Eq.(\ref{wsa}) corresponding to the given set of $\mu_{k'}$ and
$\nu_{i\neq l}$'s : 
\begin{equation}
\psi_{l}^{(l)}= C
\Hcolumn{1}{\nu_{l}^{-1}}\; , \;
\psi_{m\neq l}^{(l)}= C 
\langle m|S^{\dagger} \hat{\mu }\hat{\bbox{S}}_{l}|l \rangle 
\left(
\begin{array}{l}
\nu_{m}\\
 1
\end{array}
\right)
\;\; , \;\;
\psi_{k'}^{(l)} = C
 \langle k'| \hat{\bbox{S}}_{l}|l\rangle
\left(
\begin{array}{l}
1    \\
 \mu_{k'}
\end{array}
\right)
\;\; ;
\label{d6a}
\end{equation}
here $\nu_{l}$ is given by Eq.(\ref{b6a}),$k,m = 1,\ldots,N $, and $C$
is arbitrary.

In the same way one builds the solution where
all the $\mu $'s and $\nu$'s 
are given as input excepting $\mu_{n'}$  
adjusted to meet the condition in Eq.(\ref{05a}): 
\begin{equation}
\psi_{n'}^{(n')}= C
\Hcolumn{\mu_{n'}^{-1}}{1}
\; ,\;
\psi_{k'\neq l'}^{(n')}= C 
\langle k'|\hat{S} \hat{\nu }\hat{\bbox{S}}_{n'}^{\dagger}|n'\rangle
\left(
\begin{array}{l}
1    \\
 \mu_{k'}
\end{array}
\right)
\;\; , \;\;
\psi_{m}^{(n')}= C
\langle m|\hat{\bbox{S}}_{n'}^{\dagger}|n' \rangle 
 \left(
\begin{array}{l}
\nu_{m}\\
 1
\end{array}
\right)
\; ;
\label{f6a}
\end{equation}
$ k',m = 1,\ldots,N $,
\begin{equation}
\mu_{n'}^{-1} = 
\langle n'|\hat{S} \hat{\nu }\hat{\bbox{S}}_{n'}^{\dagger}|n'\rangle
\; .
\label{a6a}
\end{equation}
and
\begin{equation}
\hat{\bbox{S}}_{n'}^{\dagger}= \left(\hat{S} - \hat{\mu}^{(n')} \hat{S}
\hat{\nu }\right)^{-1}\; ,
\label{g6a}
\end{equation}

\section{Formal solution: II}\label{det}

Transforming $
1 -\hat{S}\hat{\nu} \hat{S}^{\dagger}\hat{\mu}
=
\left(1-\hat{S}\hat{\nu}^{(l)} \hat{S}^{\dagger}\hat{\mu} \right)
\left( 1
- \nu^{(l)}
\left(\hat{S}^{\dagger}-
\hat{\nu}^{(l)} \hat{S}^{\dagger}\hat{\mu} \right)^{-1}
 |l \rangle \langle l |\hat{S}^{\dagger}\hat{\mu} 
 \right)
 $, and using identity $\det (1 + |X \rangle \langle Y | )= 1 +
\langle Y | X \rangle$, one gets
\begin{equation}
{\cal D}(\{\nu\},\{\mu\})=
{\cal D}(\{\nu^{(l)}\},\{\mu\})
\left(1- \nu^{(l)}
\langle l|
\hat{S}^{\dagger}\hat{\mu} \hat{\bbox{S}_{l}}
|l \rangle
 \right)
\label{66a}
\end{equation}
with $\hat{\bbox{S}_{l}}$  in Eq.(\ref{c6a}),
and $\hat{\nu}^{(l)}$ denotes the matrix which differs from
$\hat{\nu}$ only in that the element $(\hat{\nu}^{(l)})_{ll}=0$.

Similarly,
\[
{\cal D}(\{\nu\},\{\mu\})=
{\cal D}(\{\nu\},\{\mu^{(n')}\})
\left(1- \mu^{(n')}
\langle n'|\hat{S} \hat{\nu }\hat{\bbox{S}}_{n'}^{\dagger}|n'\rangle
 \right)
\]
where $\hat{\bbox{S}}_{n'}^{\dagger}$ is defined in Eq.(\ref{g6a}).

Sometimes calculations become shorter 
when one changes the representation. Since the $S$-matrix in the
 Eq.(\ref{wsa}) is a scalar in the electron-hole
space, the matching  condition is unchanged 
by any rotation $\psi  \rightarrow \psi_{{\cal O}}= \hat{{\cal
O}}\psi $,
\[
\hat{{\cal O}} = {1\over 1- \mu_{0}\nu_{0}}
\left(
\begin{array}{lr}
1   & -\nu_{0}\\
 -\mu_{0}& 1
\end{array}
\right)\;,
\]
After the rotation, the basis wave function in Eq.(\ref{85a}) 
has the same form with $(...) \rightarrow (...)_{{\cal O}}$ 
\begin{eqnarray}
  \mu_{{\cal O}k'}& =  & 
{\mu_{k'}- \mu_{0}\over 1- \mu_{k'}\nu_{0}}
              \nonumber \\   
  \nu_{{\cal O}i}& =  & 
{\nu_{i}- \nu_{0}\over 1- \mu_{0}\nu_{i}}
              \nonumber \\   
A_{{\cal O}k'}& =  & A_{k'}
{1-\mu_{k'}\nu_{0}\over 1- \mu_{0}\nu_{0}}
               \nonumber\\   
B_{{\cal O}i}& =  & B_{i}
{1-\mu_{0}\nu_{i}\over 1- \mu_{0}\nu_{0}}
 \nonumber 
\end{eqnarray}
One sees that by proper rotations any pair  $\mu_{k'}, \nu_{i}$ can be
nullified in the intermediate calculations. Of course, all other
coefficients will  also be changed. Calculations done, one gets to the
original basis.

Eqs.(\ref{06a}),  and (\ref{46a}) can be written in a more compact
form. From Eq.(\ref{66a}), 
\[
{
1\over
1 - \alpha_{0l}^{(e)} \alpha_{l}^{(h)}
} = {{\cal D}_{l}\over {\cal D}_{0}}
\]
where 
\[
{\cal D}_{0} = 
{\cal D}(\{\alpha^{(e)}\}, \{\alpha^{(h)}\})
\]
and 
\[
{\cal D}_{l} = 
{\cal D}(\{\alpha^{(e)}\}, \{(\alpha^{(h)})^{(l)}\})
\]

Absorbing ${\cal D}_{l}$ into $\bbox{S}_{l}$ {\it i.e.} 
$ {\sf S}_{l}={\cal D}_{l}\bbox{S}_{l} $, 
and using obvious  $|l \rangle \beta^{(e)*} = \hat{\beta}^{(e)*} |l
\rangle $ {\it etc.}, etc 
the scattering amplitudes read
\begin{eqnarray}
   B_{l}^{(l)}& =  & {1\over {\cal D}_{0}} 
\left(
\langle l|S^{\dagger} \hat{\alpha}^{(e)} \hat{{\sf S}}_{l}|l \rangle
 - {\cal D}_{l} \alpha_{l}^{(e)} \right)
 \nonumber\\   
B_{k\neq l}^{(l)}& =  & {1\over {\cal D}_{0}}
\langle k|\hat{\beta}^{(h)}S^{\dagger} 
\hat{\alpha}^{(e)} \hat{{\sf S}}_{l}\hat{\beta}^{(e)*}|l \rangle
 \nonumber \\   
A_{k'}^{(l)} &=&
{1\over {\cal D}_{0}}
\langle k'|\hat{\beta}^{(e)} \hat{{\sf S}}_{l}\hat{\beta}^{(e)*}|l\rangle
 \nonumber 
\end{eqnarray}

\section{Transfer Matrix }\label{trans}

Another possibility for resolving the matching conditions is
via the transfer matrix $\hat{{\cal M}}_{n'\leftarrow l}$
\[
\psi_{n'}= \hat{{\cal M}}_{n'\leftarrow l}\,\psi_{l}\;\; , 
\]
 which couples the wave functions
\[
\psi_{l}= \Hcolumn{u_{l}}{v_{l}}
\;\; , \;\;  
\psi_{n'}= \Hcolumn{u_{n'}}{v_{n'}}
\]
on a selected pair of  trajectories $l$ and $n'$; 
the parameters $\mu_{k'\neq n'}$ and $\nu_{i\neq l}$ are supposed
to be given.

As usual, the transfer matrix can be built out of the elements of two
particular solutions $\Psi^{I,II}$. 
Take $\Psi^{I}$ to be the  solution 
in Eq.(\ref{d6a}) with $\mu_{n'}$ put to zero,
\newcommand{\Afl}{\langle n'| \hat{\bbox{S}}_{n'l}|l\rangle}
\newcommand{\Bl}{\langle l|\hat{S}^{\dagger}\hat{\mu}^{(n')}%
\hat{\bbox{S}_{n'l}}
|l \rangle} 
\newcommand{\Blf}{\langle l|\hat{\bbox{S}}_{n'l}^{+}|n' \rangle}
\newcommand{\Af}{\langle n'|\hat{S} \hat{\nu}^{(l)}%
\hat{\bbox{S}}_{n'l}^{+}|n'\rangle}
\[
\Psi^{I}: \;\;\;
\psi_{n'}^{(I)}= \Afl 
{1 \choose 0}
\;\; , \;\;  
\psi_{l}^{I}=
{1 \choose \Bl}
\]
 and  $\Psi^{II}$ the  solution 
Eq.(\ref{f6a}) with  $\nu_{l}=0$,
\[
\Psi^{II}: \;\;\;
\psi_{n'}^{(I)}=
{\Af
\choose 1}
\;\; , \;\;  
\psi_{l}^{(II)}=
\Blf
{0\choose 1}
\]
where
\[
\hat{\bbox{S}}_{n'l} =
\left(\hat{S}^{\dagger} - \hat{\nu}^{(l)}
\hat{S}^{\dagger}\hat{\mu}^{(n')}\right)^{-1}\;\; ,\;\;
\hat{\bbox{S}}_{n'l}^{+}= \left(\hat{S} - \hat{\mu}^{(n')} \hat{S}
\hat{\nu}^{(l)}\right)^{-1}\; .
\]
Requiring that the transfer
matrix reproduces the relations between $\psi_{n'}$ and $\psi_{l}$ in
the two solutions,  one gets the following result
\begin{equation}
\hat{{\cal M}}_{n'\leftarrow l}
=
\Blf^{-1}
\left(
\begin{array}{lr}
A& B\\
- C& 1
\end{array}
\right)
\label{i6a}
\end{equation}
where
\[
{A = \Afl \Blf -  \Bl\Af \;, 
\atop
B = \Af\;,\; C = \Bl\; .
}
\]

The determinant of the transfer matrix is
\[
{\rm Det} \left\| \hat{{\cal M}}_{n'\leftarrow l}\right\| 
= {\Afl\over\Blf }
\]

The  inverse matrix reads
\begin{equation}
\hat{{\cal M}}_{n'\leftarrow l}^{-1}
=
\Afl^{-1}
\left(
\begin{array}{lr}
1& -B\\
 C& A
\end{array}
\right)
\label{n6a}
\end{equation}

Applying the matching conditions for the conjugated waves $\ops$ in
Eq.(\ref{75a}), one can check that the the corresponding transfer matrix is
given by the inverse of that  for $\psi $ {\it i.e.} 
\[
\ops_{n'}= \ops_{l}\hat{{\cal M}}_{n'\leftarrow l}^{-1}
\]
Since $\ops_{n'}^{(1)} \psi_{n'}^{(2)}=
\ops_{l}^{(1)}\hat{{\cal
M}}_{n'\leftarrow l}^{-1}\hat{{\cal M}}_{n'\leftarrow
 l}\psi_{l}^{(2)}= \ops_{l}^{(1)}\psi_{l}^{(2)}$, the conservation law in
 Eq.(\ref{1ra}) is not affected by knots.

\subsubsection{Transfer matrix $2\times 2$ case}\label{easy}

The transfer matrix for the case when the knot mixes 2 in- to 2
out-trajectories, can be obtained from the general expression in
Eq.(\ref{i6a}).  Algebraic simplifications of rather awkward
expression gives a pretty compact result.  Here, an alternative
derivation, algebraically more transparent, is presented.

Call the trajectories of interest by 1 and 1', and consider
calculation of $\hat{{\cal M}}_{1' \leftarrow 1}$ for given $\mu_{2'}$
and $\nu_{2}$ in Eq.(\ref{85a}).  First note superconductivity
influences the transfer matrix only via the trajectories 2 and
2'. Note also that in the normal metal case when $\mu_{2'}=\nu_{2}=0$,
the transfer matrix is simply
\begin{equation}
\hat{{\cal M}}_{1' \leftarrow 1}^{(0)} =
\left(
\begin{array}{lr}
s& 0   \\
 0  & {1\over s^{*}}
\end{array}
\right)=
s\left(
\begin{array}{lr}
1& 0   \\
 0  & 0
\end{array}
\right) + {1\over s^{*}}
\left(
\begin{array}{lr}
0& 0   \\
 0  & 1
\end{array}
\right)
\label{q6a}
\end{equation}
where $s = S_{1'1}$.

Since the $S-$matrix is a electron-hole scalar, Eq.(\ref{wsa}) is
invariant relative to rotations in the electron-hole space. 
$
\hat{\psi } \rightarrow \hat{\psi }_{{\cal O}} = \hat{{\cal O}}\psi 
$, and one can  resolve the matching conditions in arbitrary basis. 

The rotation
\[
\hat{{\cal O}} = {1\over 1- \mu_{2'}\nu_{2}} 
\left(
\begin{array}{lr}
1   & -\nu_{2}\\
 -\mu_{2'}& 1
\end{array}
\right)\; , \;
\hat{{\cal O}}^{-1}=  
\left(
\begin{array}{lr}
1   & \nu_{2}\\
 \mu_{2'}& 1
\end{array}
\right)
\]
transforms ${1 \choose \mu_{2'}}$ to
${1 \choose \mu_{2'}}_{{\cal O}}= {1 \choose 0}$ and 
${\nu_{2}\choose 1}$ to ${0\choose 1}$ as if in the normal state. 
Therefore, after the rotation, the transfer matrix is given by
Eq.(\ref{q6a}), whereas
 in  the original picture
\[
\hat{{\cal M}}_{1' \leftarrow 1}= 
\hat{{\cal O}}^{-1}\hat{{\cal M}}_{1' \leftarrow 1}^{(0)}\hat{{\cal
O}} .
\]

Inserting Eq.(\ref{q6a}), the transfer matrix reads
\begin{equation}
\hat{{\cal M}}_{1' \leftarrow 1}=
{1\over 1 - \mu_{2'}\nu_{2}}
\left(
s\; \Hcolumn{1}{\mu_{2'}}
\Hrow{1}{- \nu_{2}}
-
{1\over s^{*}}\; 
\Hcolumn{\nu_{2}}{1}
\Hrow{\mu_{2'}}{- 1}
 \right).
\label{cbb}
\end{equation}
One recognizes the combinations entering $\gR_{\pm}$ in
Eq.(\ref{94a}) with the important difference that $\nu_{2}$ and
$\mu_{2'}$ are parameters of the wave functions not at the same point
but across the knot.

Finally,
\begin{equation}
\hat{{\cal M}}_{1' \leftarrow 1}=
{1+S\over 2s*} 
\left( 1 - {R\over 1+S}  
\hat{g}_{2'\bullet 2}
\right)
\label{n4a}
\end{equation}
where $S= |s|^{2}$, $R= 1- S$, and
\[
\hat{g}_{2'\bullet 2}=
{1\over 1 - \mu_{2'}\nu_{2}}
\left(
\begin{array}{lr}
1 + \mu_{2'}\nu_{2}   &  -2 \nu_{2}\\
 2 \mu_{2'}& -1 - \mu_{2'}\nu_{2}
\end{array}
\right)
\]
is normalized $\hat{g}_{2'\bullet 2}^{2}=1$, ``across-the-knot'' Green
function. It can be presented in a factorized form as follows
\begin{equation}
{1\over 2} \left(1+\hat{g}_{2'\bullet 2} \right)=
{1\over 1 - \mu_{2'}\nu_{2}}
\Hcolumn{1}{\mu_{2'}}
\Hrow{1}{- \nu_{2}}
\label{28a}
\end{equation}

Up to normalization, $\Hcolumn{1}{\mu_{2'}}= \phi_{+}^{(2')}$ where
$\phi_{+}^{(2')}$ is the knot value of $\phi_{+}$ on the trajectory
$2'$, and $\Hrow{1}{- \nu_{2}}= \ofi_{-}^{(1)}$. Taking into
consideration Eqs.(\ref{s6a}), and (\ref{a5a}), one concludes from
Eq.(\ref{28a}) that $\left(1+\hat{g}_{2'\bullet 2} \right) \propto
\left(1+\hat{g}_{2'}\right)\left(1+\hat{g}_{ 2} \right)$ so that
\begin{equation}
\left(1+\hat{g}_{2'\bullet 2} \right) = {\cal N}^{-1}
\left(1+\hat{g}_{2'}\right)\left(1+\hat{g}_{ 2} \right)
\;\; , \;\;  {\cal N}= {1\over 2}{\rm Sp}
\left(1+\hat{g}_{2'}\right)\left(1+\hat{g}_{ 2} \right)\; ,
\label{8ab}
\end{equation}
where  the
normalization ${1\over 2}{\rm Sp}\left(1+\hat{g}_{2'\bullet 2}
\right)=1$ fixes the proportionality coefficient ${\cal N}$.
This formula expresses the ``across-the-knot'' function via Green's
function on the trajectories 2 and 2'.  

After some algebra, one gets another form of Eq.(\ref{8ab}):
\[
\hat{g}_{2'\bullet 2} = {1\over 1 + 
{1\over 2}[\hat{g}_{2'},\hat{g}_{2}]_{+}}
\left(\hat{g}_{2'} + \hat{g}_{2} + {1\over
2}[\hat{g}_{2'},\hat{g}_{2}]_{-}
\right)\;\; .
\]
(The anticommutator ${1\over 2}[\hat{g}_{2'},\hat{g}_{2}]_{+}$ in the
denominator is proportional to the unit matrix and does not pose any
problem).

\begin{figure}[h]
\centerline{\epsfig{file=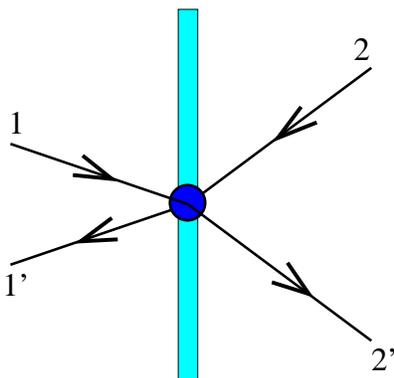,height=150pt,angle=-90}}
\caption{Scattering on a partially transparent specular interface. The
interface is depicted as
the shaded region. The arrows show the direction of the (electron) velocity. The
in-coming (out-going) trajectories are denoted 1 and 2 (1' and
2'). The filled circle, the knot (see text), is the ``black box''
where the scattering occurs. }
\label{cross}
\end{figure}

\begin{figure}[h]
\centerline{
\epsfig{file=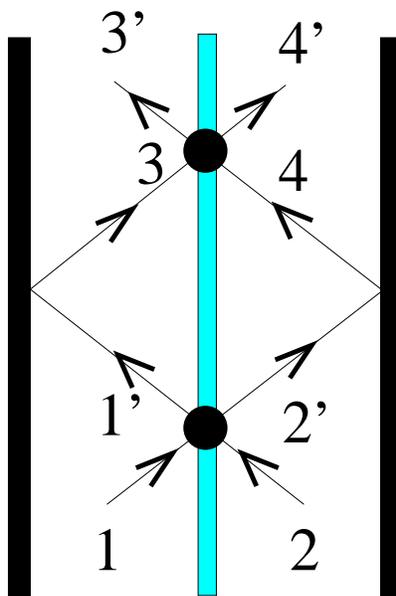,height=150pt,angle=-90}
}
\caption{The typical trajectory in an ideal sandwich with the layers of an
equal thickness and parallel surfaces. 
}
\label{loop1}
\end{figure}

\begin{figure}[h]
\centerline{
\epsfig{file=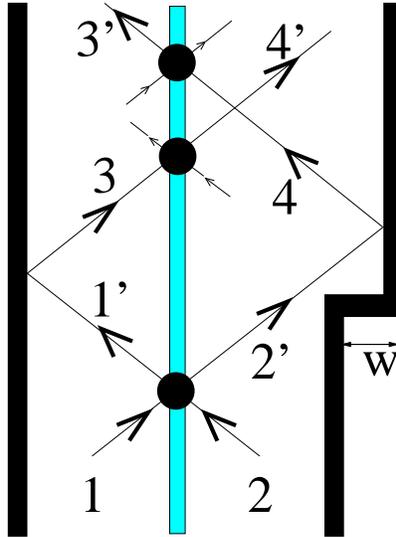,height=150pt,angle=-90}
}
\caption{The typical trajectory  in a  sandwich  with a rough surface. 
The roughness is shown schematically as a step, $W$
being the height of the step. Unlike the ideal case in
Fig. \protect\ref{loop1}, the paths 1' and 2' return to the interface at
different points.}
\label{loop2}
\end{figure}

\begin{figure}[h] 
\centerline{\epsfig{file=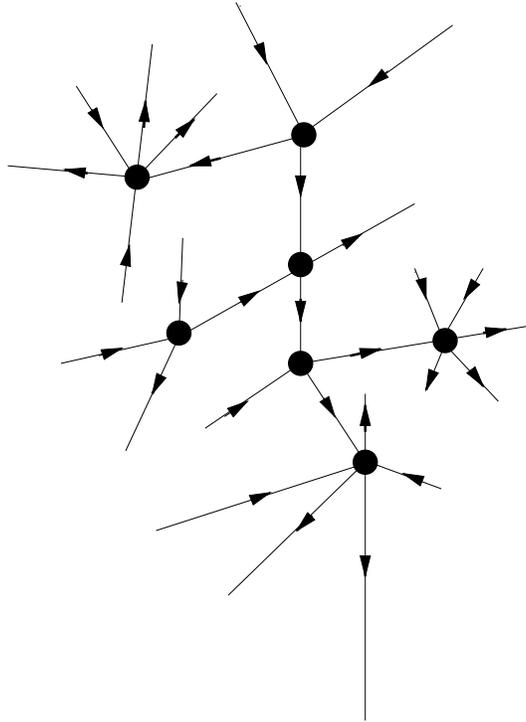,height=200pt,angle=-90}}
\caption{%
An example of a tree-like trajectory.  Pieces of the straight
lines show the trajectories before or after they enter a knot (filled
circles), 
{\it i.e.} before or after a  collision with an interface.
There is only one path connecting any two points on the tree so that
the tree is effectively 1-dimensional.
} 
\label{3-Fig} 
\end{figure}

\begin{figure}[h]
\centerline{\epsfig{file=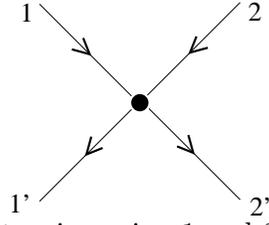,height=100pt,angle=-90}}
\caption{Simplest $2\times 2$ knot with two incoming  1 and 2,
and two outgoing channels 1' and 2' (schematically).}
\label{N=2}
\end{figure}

\begin{figure}[h] 
\centerline{\epsfig{file=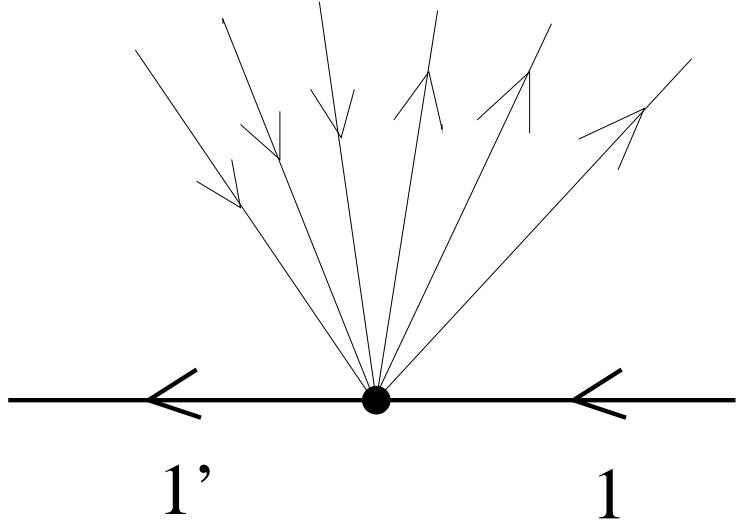,width=200pt,angle=-90}}
\caption{For a many channel knot, one chooses a pair of trajectories, one
in- and one out- (denoting them $1$ and $1'$), and considers them as
a single trajectory with a knot on it. The transfer matrix relates to
each other the wave functions across the  knot. }
\label{trans22}
\end{figure}

\begin{figure}[h]
\centerline{\epsfig{file=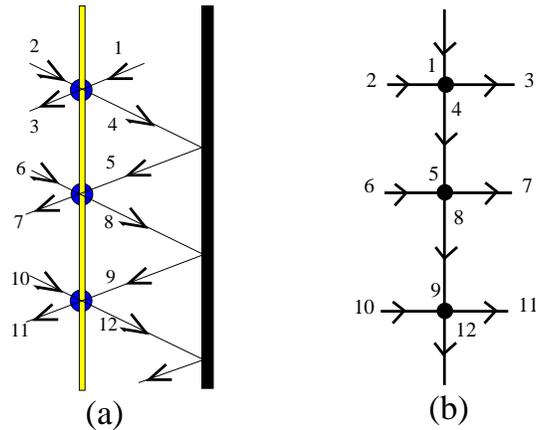,height=200pt,angle=-90}}
\caption{The typical trajectory formed by the total reflection on the
outer surface and the partial reflection/transmissions on the interface (a). The numbers
serve as markers for both  direction and position. In (b), the
structure of the tree-like trajectory is shown with the numbering as
in (a).}
\label{film}
\end{figure}

\newpage
\begin{figure}[h] 
\centerline{\epsfig{file=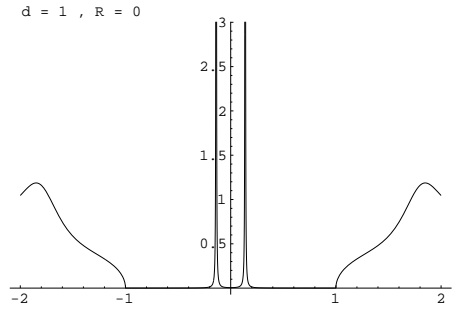,height=130pt,angle=0}}
\centerline{(a)}
\centerline{\epsfig{file=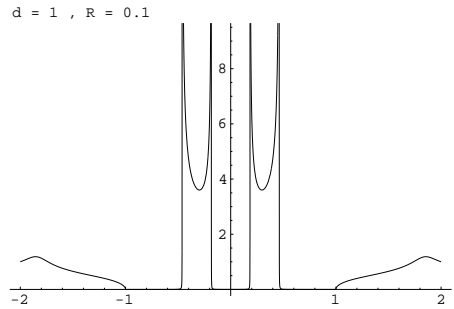,height=130pt,angle=0}}
\centerline{(b)}
\centerline{\epsfig{file=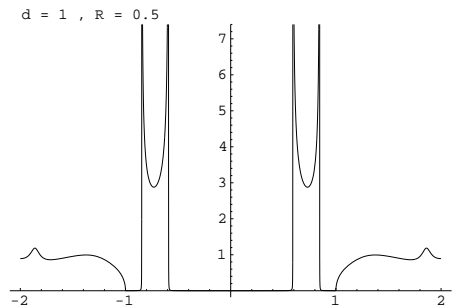,height=130pt,angle=0}}
\centerline{(c)}
\centerline{\epsfig{file=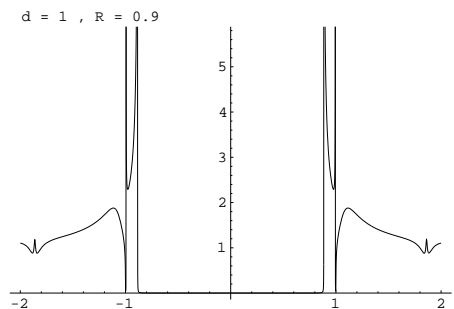,height=130pt,angle=0}}
\centerline{(d)}
\caption{The density of states (trajectory resolved) versus energy
$\varepsilon /\Delta $ at the interface of a bulk superconductor with
the pair potential $\Delta_{l}$ and a film $\Delta_{r}= - \Delta_{l}$
(see Fig. \protect\ref{film}).  The film thickness $d$ is measured
along the trajectory in units $v/\|\Delta|$. The interface
reflectivity $R=0$, 0.1, 0.5, 0.9 in (a,b,c,d) respectively.  }
\label{layer.dos}
\end{figure}
\newpage

\begin{figure}[h]
\centerline{\epsfig{file=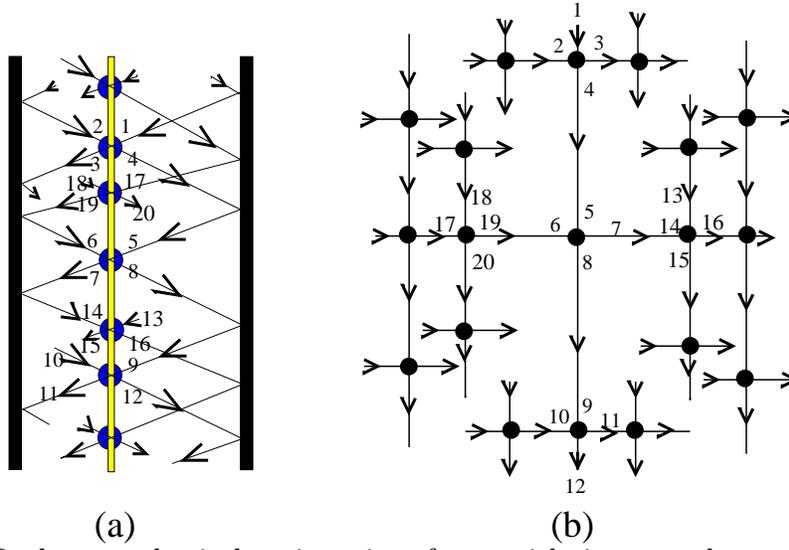,height=300pt,angle=-90}}
\caption{(a )Real space classical trajectories of a particle in a two
layers system formed by multiple reflections on the outer surface and
the interface between layers . Numbers tag both the position of the
particle on the trajectory as well as the direction.  (b) The
structure of the tree-like trajectory is shown. The points in real
space and on the tree are marked by the same numbers in (a) and (b).}
\label{sand+tree} 
\end{figure}

\newpage

\begin{figure}[h]
\noindent 
\centerline{\epsfig{file=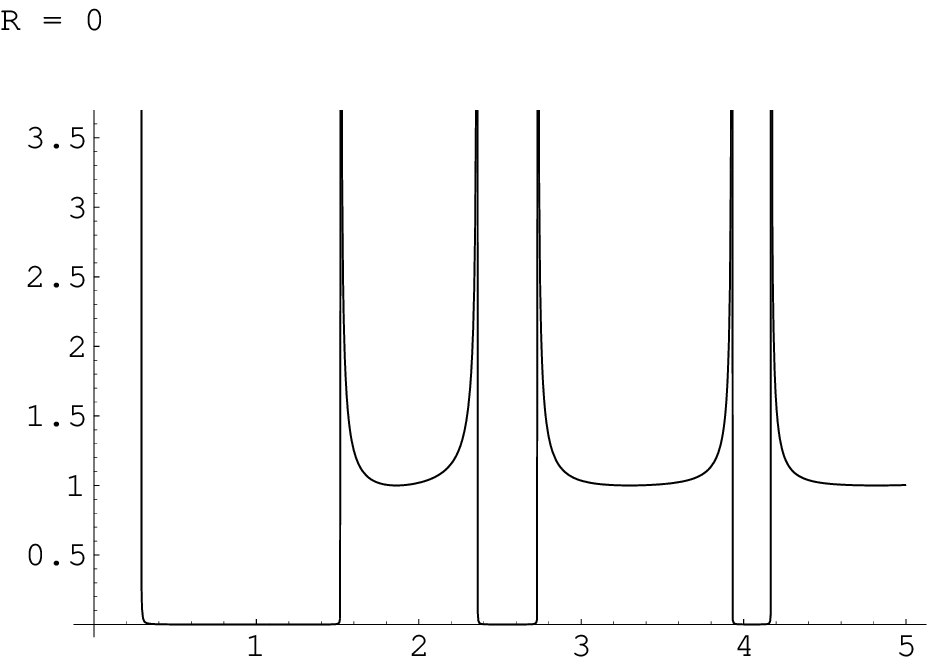,height=0.250\textwidth,angle=0}}
\centerline{(a)}

\centerline{\epsfig{file=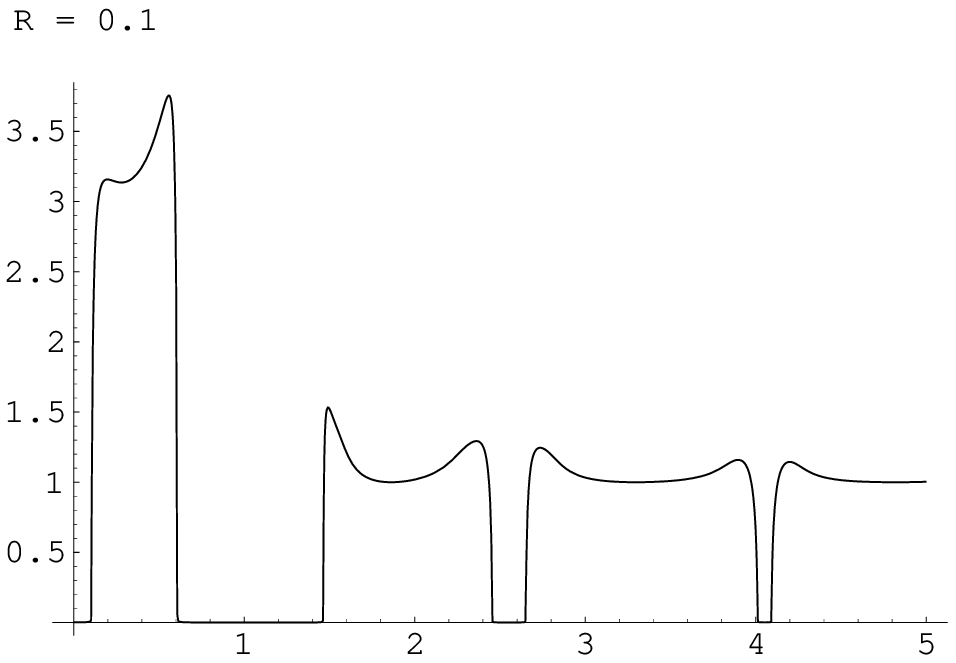,height=0.250\textwidth,angle=0}}
\centerline{(b)}

\centerline{\epsfig{file=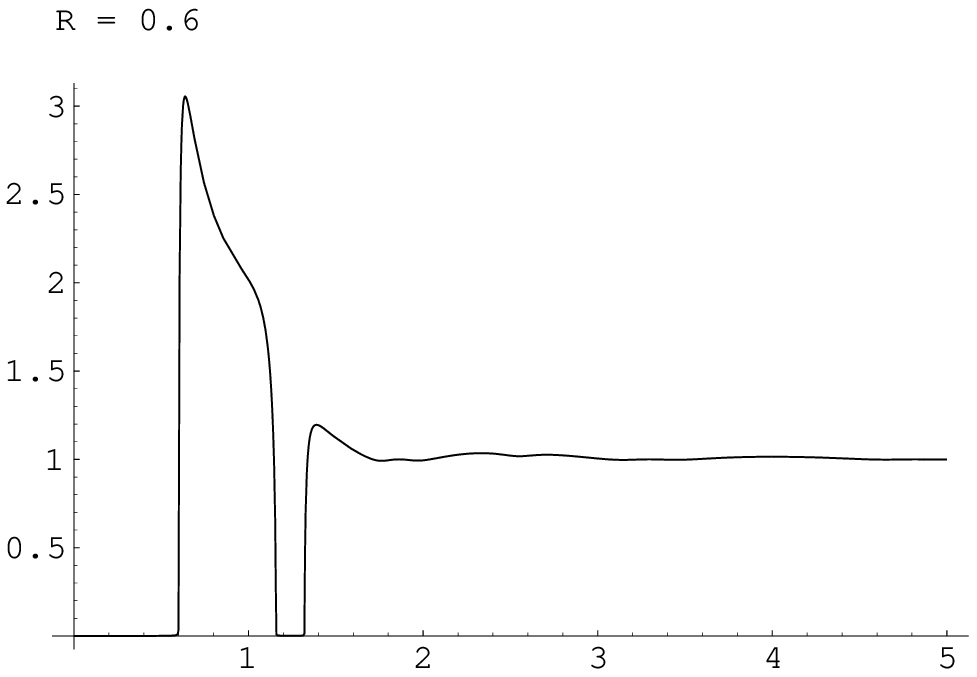,height=0.250\textwidth,angle=0}}
\centerline{(c)}

\hfill\hfill 
\centerline{\epsfig{file=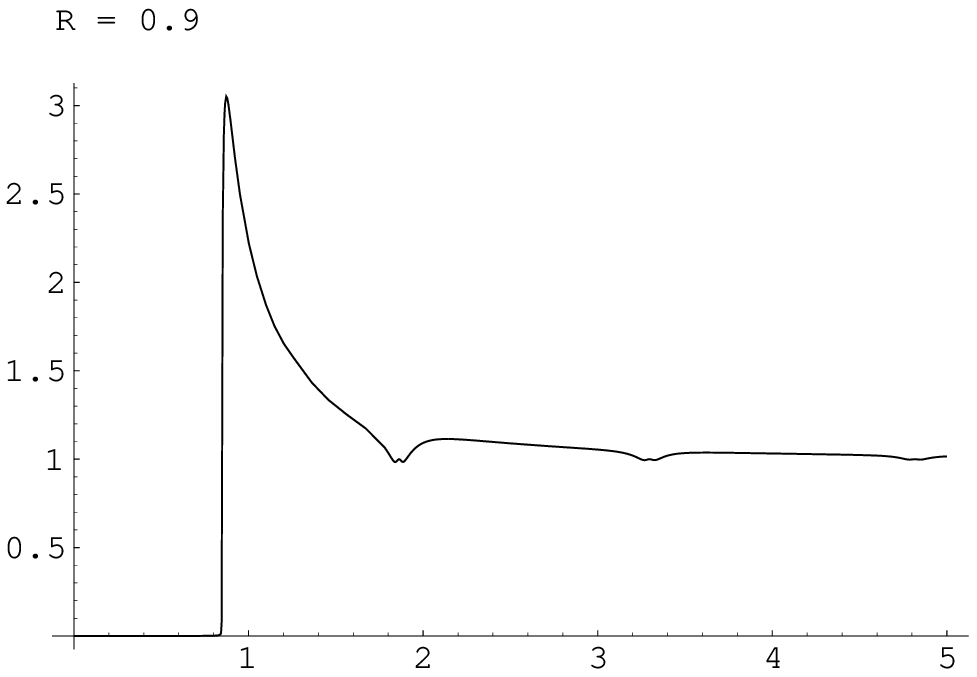,height=0.250\textwidth,angle=0}}
\centerline{(d)}
\caption{Trajectory resolved density of states
versus energy $\varepsilon/\Delta$ at the interface. The order
parameter $\Delta_{l}= - \Delta_{r}$. The thickness of the both layers
is $v/|\Delta_{l}|$. The reflection $R$, shown at the top of the
 pictures, is 0, 0.1, 0.6, and 0.9 in (a), (b), (c), and (d),
respectively. 
} 
\label{sandwich}
\end{figure}

\begin{figure}[h]
\centerline{\epsfig{file=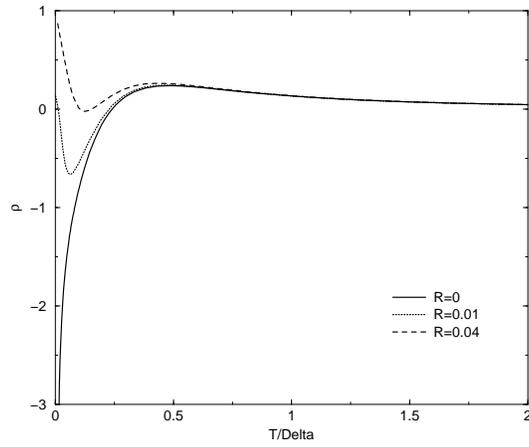,height=200pt,angle=-90}}
\caption{The effective superfluid density of a  system of two layers,
$l$ and $r$, of
equal thickness $d_{l}=d_{r}= v/|\Delta_{l}| $ with the $\pi $
phase difference $\Delta_{l}=-\Delta_{r}$ for different reflectivity 
of the interface $R=0,\,  0.01, \, 0.04$}
\label{rhoss}
\end{figure}


\begin{references}
\bibitem{e-h} It follows from Eq.(\ref{Oka}), that $\psie(\bbox{r},t
|\{\bbox{A}\})$ and $\psih^{*}(\bbox{r},t|\{\bbox{A}\}))$ satisfy same
equation.  Thus, $\psie(\bbox{r},t)=\psih^{*}(\bbox{r},t)$ provided
$\psie(\bbox{r},0)=\psih^{*}(\bbox{r},0)$.
\bibitem{traj}
Generally speaking $\bbox{\nabla }U \neq 0$ and the trajectory may be
different form a straight line. Then, one introduces the envelope $\psi
$ by the anzats $\Psi (\bbox{r},t ) = \psi (r) e^{{i\over{\hbar}}
S(\bbox{r})} e^{{i\over{\hbar}}E t}$ where $S(\bbox{r})$ is the
classical action. One gets the Andreev equation again where
$\bbox{v\cdot \nabla }$ has the meaning of the derivative 
with respect to the coordinate along the trajectory.
\bibitem{full-cross}It is important to realize that the knot couples
together all four directions.  Unlike usual scattering theory where
the in-coming channels 1 and 2 in Fig.\protect\ref{cross} would be
independent, in superconductors one must include all the channels
coupled by scattering.  The formal difference arises due to the
two-component character of the wave functions. In physical terms, the
arrows show the direction of the momentum rather than velocity; apart
the momentum, the latter depends also on the character of an
excitation, electron or hole.  On the knot, an electron-like
excitation coming along 1 may be converted
into a hole which goes away along nominally in-coming trajectory 2.
\bibitem{curve}
The phase of the exponents in Eq.(\ref{qza}) is the free motion
classical action.  If the particle is subject to a smooth
time-reversal symmetry conserving potential, it can be easily
incorporated into the scheme. In this case the trajectory is found
from the classical equation of motion.
\bibitem{traceless} 
More generally, the matrix $\hat{H}^{R}$ has the contribution $(-\Phi)
\hat{1}$ where $\Phi = e \varphi + \dot{\chi}/2$ is the gauge
invariant potential made of the scalar potential $\varphi $ and the
phase of the order parameter $\chi$. The substitution $g \rightarrow
\exp{\Big(i\pm \int\limits_{x_{1}}^{x_{2}} dx\ \Phi (x)
\Big)}\hat{g}(x_{1},x_{2})$ eliminates $\Phi $ from the equations.
\bibitem{irred} The knot assumed to be ``irreducible'' {\it i.e.}  the
channels can not be combined in disconnected groups with the number of
channels less than $N$.
\bibitem{real}Note that the amplitude of the Andreev reflection by the knot, $B_{l}$, is a real function of $a$'s and $b$'s , {\it i.e.} 
the coefficients in the expansion of $B_{l}$ in series of
$a_{i}^{n}b_{f}^{m}$,
which are combinations of the S-matrix elements, are real.
\end{references}
\end{document}